\documentclass[useAMS,usenatbib]{mn2e}     
\usepackage{epsfig}

\def\spose#1{\hbox to 0pt{#1\hss}}        
\def\lta{\mathrel{\spose{\lower 3pt\hbox{$\mathchar"218$}}        
     \raise 2.0pt\hbox{$\mathchar"13C$}}}        
\def\gta{\mathrel{\spose{\lower 3pt\hbox{$\mathchar"218$}}        
     \raise 2.0pt\hbox{$\mathchar"13E$}}}        
     
\title[Star formation feedback and metal enrichment in dSphs]{Star 
formation feedback and metal enrichment by SN Ia and SN II in dwarf 
spheroidal galaxies: the case of Draco} 
 
\author[A. Marcolini, A. D'Ercole, F. Brighenti and S. Recchi]   
       {A. Marcolini$^{1,2}$, A. D'Ercole$^2$, F. Brighenti$^1$ and  
         S. Recchi$^3$  \\   
        $^1$ Dipartimento di Astronomia, Universit\`a di Bologna,   
        via Ranzani 1, 40127 Bologna, Italy \\   
        $^2$ Osservatorio Astronomico di Bologna,   
        via Ranzani 1, 40127 Bologna, Italy \\   
        $^3$ Institute of Astronomy, Vienna University,  
        T\"urkenschanzstrasse 17, 1180 Vienna, Austria} 
 
\date{Accepted ..., Received ...; in original ...}   
   
\pagerange{\pageref{firstpage}--\pageref{lastpage}}   
   
\pubyear{2004}   
    
\begin{document}   
   
\maketitle   
   
\label{firstpage}   
   
\begin{abstract}  
   
We present 3D hydrodynamic simulations aimed at studying the dynamical 
and chemical evolution of the interstellar medium in dwarf spheroidal 
galaxies. This evolution is driven by the explosions of Type II and 
Type Ia supernovae, whose different contribution is explicitly taken 
into account in our models. We compare our results with detailed 
observations of the Draco galaxy. We assume star formation histories 
consisting of a number of instantaneous bursts separated by quiescent 
periods. Diverse histories differ by the number of bursts, but all 
have the same total duration and give rise to the same amount of 
stars. Because of the large effectiveness of the radiative losses and 
the extended dark matter halo, no galactic wind develops, despite the 
total energy released by the supernovae is much larger than the 
binding energy of the gas.  This explains why the galaxy is able to 
form stars for a long period ($> 3$ Gyr), consistently with 
observations. In this picture, the end of the star formation and gas 
removal must result from external mechanisms, such as ram pressure 
and/or tidal interaction with the Galaxy. The stellar [Fe/H] 
distributions found in our models match very well  
the observed ones. We find a mean value $\langle$[Fe/H]$\rangle = 
-1.65$ with a spread of $\sim 1.5$ dex. The chemical properties of the 
stars derive by the different temporal evolution between Type Ia and 
Type II supernova rate, and by the different mixing of the metals 
produced by the two types of supernovae. We reproduce successfully the 
observed [O/Fe]-[Fe/H] diagram.  However, our interpretation of this 
diagram differs from that generally adopted by previous chemical 
models. In fact, we find that the break observed in the diagram is not 
connected with the onset of a galactic wind or with a characteristic time 
scale for the sudden switchover of the Type Ia supernovae, as sometimes 
claimed.  Instead, we find that the chemical properties of the stars 
derive, besides the different temporal evolution of the SNe II and SNe Ia 
rates, from the spatial inhomogeneous chemical enrichment due to the 
different dynamical behaviour between the remnants of the two types of  
supernovae. 
 
\end{abstract}  
  
   
\section{Introduction}

Dwarf galaxies are the most common class of galaxies in the local  
Universe \citep{marzke1997} and they probably were much more numerous  
at past epochs \citep{ellis1997}. Dwarf spheroidal/elliptical galaxies  
(dSph/dE) are found preferentially in high density environments and can  
be studied in nearby clusters \citep{popesso2006} like Coma  
\citep{secker1997}, Fornax \citep{rakos2001}, Virgo  
\citep{binggeli1991, phillipps1998} and Perseus \citep{conselice2003}.  
However, due to their proximity, dSphs of the Local Group (see Mateo  
1998 for a review) offer an unique opportunity to study in detail  
their structural properties, formation and chemical evolution.  
  
Dwarf spheroidals are the least massive galaxies known, but yet, their 
velocity dispersions imply mass to light ratios as large as 100 
$\rm M_{\odot}/L_{\odot}$. This is usually explained assuming that these 
systems are dark matter dominated. Actually, in the past few years 
both observational evidences \citep[e.g.][]{lokas2002, wilkinson2002, 
kleyna2002, walker2005} and theoretical works \citep{kazantzidis2004, 
mayer2005, mashchenko2005, mashchenko2006} confirm the possibility 
that these galaxies are relatively massive bounded system with virial 
masses in the range $10^8-5\times10^9$ M$_{\odot}$; but, for a 
dissonant view see \citet{kroupa1997}, \citet{kroupa2005}, or the review  
by Gallagher \& Wyse 1994.

Such galaxies are very metal poor and lack of neutral hydrogen and 
recent star formation. Thus they were initially believed to be very 
similar to Galactic globular clusters and to have a very simple star 
formation history (SFH). Recent studies have shown, instead, that 
these systems are much more complex, with varied and extended SFHs. 
High resolution spectroscopy of several dSphs showed the presence of a 
wide range in metallicity \citep{harbeck2001}.  For example, abundance 
analyses of stars belonging to Draco and Ursa Minor have shown 
values of [Fe/H] in the range $-3 \leq$[Fe/H]$\leq -1.5$ 
\citep{stetson1984, shetrone1998, shetrone2001} with a mean 
value in the interval $-2.0 \leq \langle$[Fe/H]$\rangle 
\leq -1.6$, depending on the authors \citep{aparicio2001, 
shetrone2001, bellazzini2002}. 
 
The above ranges are consistent, for some dSphs, with a 
single period of star formation extended in time for a few Gyr 
\citep{carney1986, mateo98,dolphin2002, babusiaux2005}. As a further 
hint of long SFH \citet{shetrone2001} found that their observed dSphs 
have [$\alpha$/Fe] abundances that are $\sim 0.2$ dex lower than those 
of Galactic halo field stars in the same [Fe/H] range. This 
suggests that the stars in these systems were formed in gas 
pre-enriched by Type II supernovae (SNe II) as well as by Type Ia 
supernovae (SNe Ia), and star formation must thus continue over a 
relatively long timescale in order to allow a sufficient production of 
iron by SNe Ia (typically $> 1-2$ Gyr). 
  
Given the small dynamical mass inferred for dSphs, the interstellar 
medium (ISM) binding energy is small when compared to the energy 
released by the SNe II explosions occurring during the star formation 
period; for instance, as shown in the next section, in a dSph as 
Draco, the baryonic matter has a binding energy of $\sim 10^{53}$ erg, 
while the expected number of SN explosions in the past was 
$10^3-10^4$, realising an energy much larger than the binding 
energy. It is thus quite puzzling how the ISM can remain bound long 
enough to allow such a long star formation duration. Actually, 
cosmological simulations \citep[e.g.][here after RG]{kawata2005, 
  ricotti2005} find rather short durations ($< 1$ Gyr) of the star 
formation, and several authors believe that the ISM of these systems 
may be completely removed by violent SNe explosions \citep{dekel1986, 
  mori1997, murakami1999, mori2002, mori2004, hensler2004} together 
with the newly synthesized metals. 
  
Gas removal via galactic winds powered by SN explosions is also  
invoked by \citet{lanfranchi2004}(hereafter LM) who proposed a one  
zone chemical evolution model with very low star formation  
efficiencies and high wind efficiency, which is able to reproduce the  
metallicity distribution and the general features of these galaxies.  
On the other hand, such features may be obtained also by closed box  
chemical evolution models by  \citet{ikuta2002}(here after IA) who 
prescribe a low star formation rate (SFR) and relatively long duration  
of the SFH, as suggested by observations.  
  
Motivated by the above arguments, in this paper we explore the 
possibility that dSphs galaxies formed stars at a low SFR for a long 
period.  To compare our results with observations, we have tailored 
our models on the Draco galaxy: this galaxy is supposed to have 
experienced a star formation lasting for 3-4 Gyr, and which 
essentially ceased 10 Gyr ago \citep{mateo98}. Obviously, our results 
may be confronted with other dSphs which are strongly dark matter 
dominated and have similar star formation histories as, e.g., Ursa Minor 
\citep{mateo98}. 
 
We run a number of three-dimensional (3D) hydrodynamical 
simulations to study the dynamical and chemical evolution of this 
system, following an assumed star formation history. A special 
attention is paid to the influence of both SNe Ia and SNe II on the 
chemical enrichment of the new forming stars.


\section{The model}  
\label{sec:model} 
 
In this section we describe as we build up our model.  We start our 
simulations with the ISM in hydrostatic equilibrium in the dark matter 
halo potential well. Any baryonic (stars or gas) contribution to the 
gravitational potential is neglected. We approximate the Draco 
observed stellar distribution with a King profile in order to properly 
locate the forming stars and the SNe explosions on the numerical grid 
(see Section~\ref{sec:supernovae_explosions}). The dark halo properties 
are derived by the observed mass to light ratio, as described below. 
  
\subsection{Galactic model}   
   
We assume a quasi-isothermal dark halo, whose density is  
given by:  
   
\begin{equation}   
\rho_{\rm h}(r)=\frac {\rho_{\rm h,0}} {1+\left (\frac{r}{r _{\rm h,c}}   
\right )^2},   
\label{equ:halo_dark}    
\end{equation}   
where $r _{\rm h,c}$ is the core radius.    
The halo mass as a fuction of radius is then:   
\begin{equation}   
M_{\rm h}(r)=4 \pi \rho_{\rm h,0} r_{\rm h,c}^3 (y - {\arctan} y),   
\label{equ:halo_mass}    
\end{equation}   
where $y=r/r_{\rm h,c}$. The density profile is truncated at a tidal  
radius $r_{\rm h,t}$ in order to obtain a    
finite mass. The gravitational potential given by this mass profile is:   
   
\begin{equation}   
\Phi_{\rm h}(r)=4 \pi G \rho_{\rm h,0} r_{\rm h,c}^2 \left[\frac{1}{2}    
\log (1+y^2) + \frac{\arctan y}{y}\right].   
\label{equ:halo_pot}   
\end{equation}

\begin{table*} 
\centering   
\begin{minipage}{170mm}   
\caption{Galaxy parameters} 
\label{tab:galaxy}   
\begin{tabular} {|c|c|c|c|c|c|c|c|c|c|c|c|c|c|}   
\hline   
Model &  $M_{\rm h}  $ & $\rho_{\rm h,0}$ &      
         $r_{\rm h,c}$ & $r_{\rm h,t} $ &   
         $M_{\star}$ & $\rho_{\star, \rm 0}^a$ &   
         $r_{\rm eff}^b$  & $r_{\star, \rm c} $ &  $r_{\star, \rm t}^c$ &   
         $(M/L_{\rm V})_0^d$  & $(M/L_{\rm V})_{\rm tot}$  \\

         & ($10^6$ M$_{\odot}$) & ($10^{-24}$ g cm$^{-3}$) & (pc)  & (pc)      
         & ($10^6$ M$_{\odot}$) & ($10^{-24}$ g cm$^{-3}$) & (pc)  & (pc) &  
(pc)   
         & ($\rm M_{\odot}/L_{V,\odot}$) &  ($\rm M_{\odot}/L_{V,\odot}$) \\  
  
\hline   
Draco   & 62 & 4.3 & 300 & 1222 & 0.56 & 1.0 & 210 & 130 & 650 & 80 & 211 \\  
Draco S & 22 & 6.5 & 160 & 1005 & 0.56 & 1.0 & 210 & 130 & 650 & 14 & 80  \\  
\hline  
\end{tabular}   
\par  
\medskip 
All the above quantities are derived from the models, but:   
$^a$ the observed central stellar density \citep{mateo98},  
$^b$ the observed effective radius of the stellar component \citep{peterson93},
$^c$ the observed tidal radius \citep{irwin1995}. 
$^d$ Value calculated inside the stellar volume.  
\end{minipage}   
\end{table*}   
  
We assume the following relations among the dark halo parameters: 
   
\begin{equation}   
\rho_{\rm h,0}=6.3 \times 10^{10} \left (\frac{M_{\rm h}}{\rm M_\odot} 
\right)^{-1/3} h^{-1/3} {\rm M_{\odot}} {\rm kpc^{-3}}, 
\label{equ:halo_rho}    
\end{equation}   
   
\begin{equation}   
r_{\rm h,c}=8.9 \times 10^{-6} \left (\frac{M_{\rm h}}{\rm M_\odot} 
\right)^{1/2} h^{1/2} {\rm kpc}, 
\label{equ:halo_rcore}    
\end{equation}   
   
\begin{equation}   
r_{\rm h,t}=0.016 \left (\frac{M_{\rm h}}{\rm M_\odot} \right)^{1/3} 
h^{-2/3} {\rm kpc}, 
\label{equ:halo_rcut}   
\end{equation}   
\noindent   
where $h=0.7$ is the assumed value of the normalized Hubble constant. 
The above equations are obtained by \citet {maclow99} in the case of a 
quasi-isothermal model on the basis of similar relations found for dark 
matter halos with Burkert profiles in the range 
$4 \times 10^9 < M_{\rm h} < 10^{11}$ M$_{\odot}$ \citep{burkert95}. 
Following \citet{maclow99} and \citet{silich01} 
we scale down these relations to halos with lower masses.

In the present paper we tailor our models on the Draco galaxy.  The 
values of the parameters in the above formulae are constrained by 
observations of the mass to light ratio of this galaxy. We thus take 
into account the stellar component assuming a density profile given by 
the King profile: 
  
\begin{equation}   
\rho_{\star}(r)=\frac {\rho_{\star, \rm 0}} {\left (1+ \left (\frac{r} 
{r _{\star, \rm c}}\right)^2 \right )^{3/2}}, 
\label{equ:star_density}    
\end{equation}   
where $r _{\star, \rm c}$ is the stellar core radius.  The stellar 
mass inside a given radius $r$ is: 
\begin{equation}   
M_{\star}(r) =4 \pi \rho_{\star, \rm 0} r_{\star, \rm c}^3 \left (\ln 
(x-\sqrt{1+x^2})- \frac{x}{\sqrt{1+x^2}} \right) 
\label{equ:star_mass}    
\end{equation}   
where $x=r/r_{\star, \rm c}$. As for the dark halo, the stellar 
profile must be truncated at a tidal radius $r_{\star, \rm t}$ in 
order to obtain a finite stellar mass. 
In the following we will refer to the volume inside $r_{\star, \rm 
t}$ as the ``stellar region'', and to the volume inside $r_{\rm h, t}$ 
as the ``galactic region''. 
 
From observations of the central stellar luminosity density and the 
total V-band luminosity \citep[e.g.][]{mateo98}, and adopting a 
stellar mass to light ratio in the V band $M_{\star}/L_{\rm V}=2 
\rm M_{\odot}/L_{\odot}$, we estimate that $\rho_{\star,0}= 1.0 \times 
10^{-24}$ g cm$^{-3}$ and $M_{\star}=M_{\star}(r_{\star, \rm t})=5.6 
\times 10^5$ M$_{\odot}$ \citep[e.g.][]{mateo98}.  We choose 
$r_{\star,\rm c}= 130$ pc and $r_{\star,\rm t}=650$ pc \citep[see 
e.g.][]{lake1990, pryor1990, irwin1995, piatek2001, mateo98} which 
give the values of $M_{\star}$ estimated above and also an effective 
radius $r_{\rm eff}=220$ pc, in good agreement with the observed value 
$r_{\rm eff}=210$ pc \citep{peterson93}.  Assuming that the total mass 
to light ratio in the volume up to $r_{\star, \rm t}$ is $(M_h(\rm 
r_{\star, \rm t})+M_{\star})/L_{\rm V}=80 \; \rm M_{\odot}/L_{\rm 
V,\odot}$ \citep{mateo98}, we obtain the dark matter mass inside the 
stellar volume. In order to obtain the total dark halo mass $M_{\rm 
h}=M_{\rm h}(r_{\rm h,t})$ we solve iteratively 
equations~\ref{equ:halo_rho}, \ref{equ:halo_rcore} and 
\ref{equ:halo_rcut} in order to find a dark halo with the right value 
of $M_h(\rm r_{\star, \rm t})$. Only one value of $M_{\rm h}$ exists, 
which gives the observed mass to light ratio. This value turns out 
to be $M_{\rm h}= 1.7 \times 10^9$ $M_{\odot}$, in agreement with most 
recent inferred halo masses $10^8-5 \times 10^9$ M$_{\odot}$ 
\citep[see e.g.][]{kleyna2002, kazantzidis2004, wilkinson2004, 
lokas2002, walker2005, mashchenko2005}.

\begin{table*}   
\centering   
\begin{minipage}{100mm}   
\caption{ISM parameters}   
\label{tab:ism} 
\begin{tabular} {|c|c|c|c|c|c|}   
\hline   
Model & $\rho_{\rm ISM,0}$ &  $T_{\rm ISM}$  &   
        $M_{\rm ISM, \star}^a$  &  $M_{\rm ISM,h}^b$ &   
        $E_{\rm bind}^c$ \\   
           
        & ($10^{-24}$ g cm$^{-3}$) &  (K) &   
        ($10^6$ M$_{\odot}$) & ($10^6$ M$_{\odot}$) & ($10^{51}$ erg)\\   
\hline  
Draco   & 0.4  & $1.0 \times 10^4$ &   3.3 & 11.3  & 83 \\   
Draco S & 2.3  & $2.0 \times 10^3$ &   1.7 &  3.8  & 11 \\   
\hline   
\end{tabular}   
\par  
\medskip  
$^a$ Amount of gas inside the stellar volume;  
$^b$ Amount of gas inside the dark halo volume;  
$^c$ Gas binding energy.  
\end{minipage}   
\end{table*}   
  
We take into account the effect of the tidal interaction due to the 
Milky Way considering a somewhat less extended dark halo than that 
indicated by equation~\ref{equ:halo_rcut} \citep[see e.g.][]{read2005, 
  mayer2005, mashchenko2006}. Thus we truncate the dark matter profile 
at $r_{\rm h,t}=1.2$ kpc (instead of 24.2 kpc) reducing the dark halo  
mass to a value of $M_{\rm h}=6.2 \times 10^7$ M$_{\odot}$. This truncation  
does not 
affect the stellar component which is located in the inner region of 
the dark halo, and thus is much less affected by the tidal interaction 
\citep[e.g.][]{gao2004}. The complete list of parameters concerning the 
galactic model is given in Table~\ref{tab:galaxy}. 
  
In this Table a model called Draco-S is also listed.  This model has 
been considered to test the possibility that the dark halo could be 
lighter than that assumed above. Actually, some authors claim that 
dSphs could be not virialized \citep[see e.g.][]{gallagher1994, 
kroupa1997}. In this case the dark halo would be absent or with a mass 
similar to that of the stellar component. We did not consider such an 
extreme case but, in order to explore the effect of a lighter dark 
matter halo on our models, we run separated simulations in which 
$M_{\rm h}= 2.2 \times 10^7$ M$_{\odot}$ which gives a $total$ 
$M/L_{\rm V}=80$ $\rm M_{\odot}/L_{\odot}$. As a consequence the dark halo 
mass distribution is more concentrated (cf. Table~\ref{tab:galaxy}), 
while the stellar component is the same of the more massive models. 
  
We place the ISM in isothermal equilibrium within the potential well, 
with a temperature of $T_{\rm ISM} \sim T_{\rm vir}$. The initial gas 
mass is $M_{\rm ISM} = 0.18 M_{\rm h}$, which corresponds to a 
baryonic fraction given by \citet{spergel2006}. 
The value of the central gas density $\rho_{\rm ISM,0}$ is then fixed 
by the assumption that the ISM extends up to the dark matter 
component.  The complete list of parameters concerning the ISM is 
given in Table~\ref{tab:ism}. 
  
Finally, for computational reasons, we assume that the ISM has an 
initial metallicity $Z_{\rm ISM}=10^{-4} \rm Z_{\odot}$. 
  
\subsection{Supernovae explosions}  
\label{sec:supernovae_explosions} 
 
In order to understand the response of the ISM to the SFH and the SFR, 
we changed both of them in a number of models all with the same 
initial conditions. We assumed that the star formation starts when the 
galaxy has accreted all its baryonic mass and that stars form in a 
sequence of $N_{\rm burst}$ instantaneous bursts, separated by 
quiescent periods. Several $N_{\rm burst}$ have been considered (see 
Table~\ref{tab:sne}) all giving rise to a final amount of stars equal 
to $M_{\star}$ after $t_{\rm f}=3$ Gyr, which is the end of the 
simulation. Obviously the lower is $N_{\rm burst}$ the higher is the 
intensity of each single burst (cf. Table~\ref{tab:sne}). The 
prolonged star formation history adopted is consistent with recent 
SFHs inferred for this galaxy by observations \citep[see 
e.g.][]{carney1986, mateo98, dolphin2002} and other theoretical works 
\citep[e.g.][]{ikuta2002}. 
  
After each starburst the stellar mass is increased by an amount 
$\Delta M_{\star}=M_{\star}/N_{\rm burst}$ which is distributed in 
space following equation~\ref{equ:star_density}. We assume that the 
density profile of the stars does not change with time and that the 
newly forming stars are distributed according to the same nowadays shape; such 
an assumption is supported by several studies \citep{mashchenko2005, 
kawata2005}.

\begin{figure}   
\begin{center}   
\psfig{figure=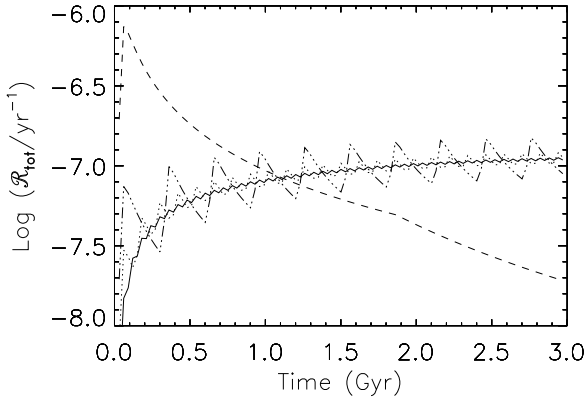 ,width=0.48\textwidth}  
\end{center}   
\caption{Logarithm of the total SN Ia rate ${\cal R_{\rm SNIa}^{\rm 
tot}}$ for four different star formation histories, all producing the 
same final mass of stars $M_{\star}=5.6 \times 10^5$ M$_{\odot}$ after 
$t_{\rm f}=3$ Gyr. Solid line: 50 bursts; dotted line: 25 bursts; dot 
dashed line: 10 bursts; dashed line: single burst.} 
\label{fig:snirate}   
\end{figure}   
  
The SNe II explode at a constant rate ${\cal R}_{\rm SN II}$ for a 
period of $\tau=30$ Myr (the lifetime of a 8 M$_{\odot}$ star, the 
least massive SN II progenitor) after each stellar burst.  Thus the SN 
II rate for each burst is given by: ${\cal R}_{\rm SN II}=N_{\rm SN 
II}/\tau,$ where $N_{\rm SN II}$ is the total number of SN II 
explosions occurring in a single burst. For a Salpeter initial mass 
function (IMF) it is $N_{\rm SN II}=0.01 \times (\Delta 
M_{\star}/ \rm M_{\odot})$.  Obviously $\Delta M_{\star}$, $M_{\star}$ and 
$N_{\rm burst}$ satisfy the simple relation $M_{\star}=N_{\rm burst} 
\Delta M_{\star}$. Our SFH also satisfies the relation $t_{\rm 
f}=N_{\rm burst}\Delta t_{\rm burst}$, where $\Delta t_{\rm 
burst}$ is the time interval between two successive 
bursts. Table~\ref{tab:sne} summarizes the models characterized by 
different choices of the above parameters. 
   
The SN Ia rate ${\cal R}_{\rm SN Ia}$ for a single population 
decreases in time after an initial rise \citep[e.g.][]{greggio83, 
  matteucci2001, greggio2005}.  We adopted the time dependent rate 
given by \citet{matteucci2001} according to the Single-Degenerate 
scenario (equation 2 of their paper with a fraction of binary systems 
$A=0.006$).  For every burst the time $t_{\rm i}$ at which each SN Ia 
explodes is given by the condition: 
$$\int_{\rm t_{i-1}}^{\rm t_i} {\cal R}_{\rm SN Ia}(t) dt=1,$$  
where $t_{\rm i-1}$ is the time at which the previous SN Ia (belonging  
to the same burst) exploded.

\begin{figure} 
\begin{center}  
\psfig{figure=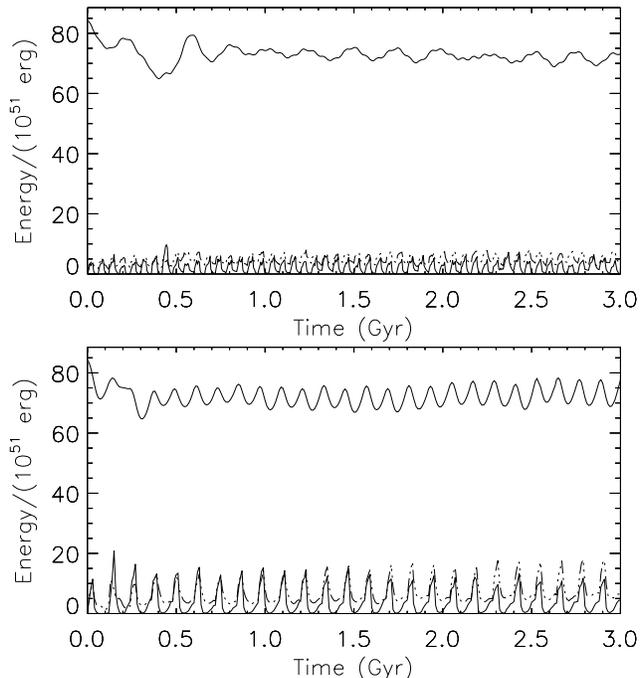,width=0.5\textwidth} 
\caption{Time evolution of the binding (solid line), thermal (thin 
solid line) and kinetic (dot dashed line) energies of the ISM. The 
upper panel refers to the reference model Draco-50, while the lower 
panel refers to the model Draco-25.} 
\label{fig:ebild}  
\end{center}   
\end{figure}

In Fig.~\ref{fig:snirate} we plot the time profile of the total SN Ia 
rate ${\cal R}_{\rm SN Ia}^{\rm tot}$ (given by the sum of all ${\cal 
R}_{\rm SN Ia}$ due to the bursts occurred within that time) for four 
different histories of star formation: a single burst, 10 bursts, 25 
bursts and 50 bursts, all producing the same amount of total stellar 
mass $M_{\star}$ after a time $t_{\rm f}$. For a single instantaneous 
burst the SNe Ia start occurring several Myr ($>30$ Myr) after the 
onset of the star formation and continue for several Gyr owing to the 
long lifetimes of the stars responsible of this type of 
explosions. Note that when more than one burst is present ${\cal 
R}_{\rm SNIa}^{\rm tot}$ has an oscillatory behaviour overimposed on 
the general increase with time.  In contrast with ${\cal R}_{\rm SN 
II}$, ${\cal R}_{\rm SNIa}^{\rm tot}$ does not have an intermittent 
nature.  This difference, together with the different chemical 
composition of the ejecta of the two types of supernovae (see below), 
is responsible of some chemical properties of the stellar population 
(see Section~\ref{sec:chemical_evolution}). 
   
Each SN explosion is stochastically placed into the galaxy. The 
probability of a SN explosion inside a radius $r$ is proportional to 
the stellar mass within that radius: $ P(r)=M_{\star}(r)/M_{\star}$. 
The angular distribution is obtained choosing randomly the azimuthal 
angle $\Phi$ and the cosine of the polar angle $\theta$. 
  
\begin{table*}   
\centering   
\begin{minipage}{100mm}   
\caption{Supernovae parameters}  
\label{tab:sne} 
\begin{tabular} {|c|c|c|c|c|c|}  
\hline  
Model & $N_{\rm burst}$   
      & $N^a_{\rm SN II}$  
      & $(N_{\rm SN II}/N_{\rm burst})^b$  
 & $ \Delta t^c_{\rm burst} $ & $N^d_{\rm SNIa}$ \\  
  
       &  &  &  & (Myr) & \\  
  
\hline  
  
Draco (S)-50  & 50 & $5.6 \times 10^3$ & 112 &  60 &  254 \\  
Draco (S)-25  & 25 & $5.6 \times 10^3$ & 224 & 120 &  257 \\  
Draco (S)-10  & 10 & $5.6 \times 10^3$ & 560 & 300 &  266 \\   
  
\hline  
  
\end{tabular}  
\par  
\medskip  
$^a$ Total number of SNe II;  
$^b$ Number of SNe II forming in each stellar burst;  
$^c$ Quiescent period between two consecutive stellar bursts;  
$^d$ Total number of SNe Ia after 3 Gyr. 
\end{minipage}   
\end{table*}

Finally, we assume that each SN II ejects a mean mass of $M_{\rm SN
II, ej}=10 \; \rm M_{\odot}$, and each SN Ia ejects $M_{\rm SNIa, ej}=1.4
\; \rm M_{\odot}$.  Every SN II expels 1.0 M$_{\odot}$ of oxygen and $0.07$
M$_{\odot}$ of iron \citep[consistent with case A of ][and reference
therein]{woosley95, gibson97b} while each SN Ia ejects 0.15
M$_{\odot}$ of oxygen and 0.74 M$_{\odot}$ of iron \citep[and
reference therein]{gibson97b}. These yields slightly depend on
metallicity, unless metals are completely absent \citep{woosley95}; we
do not consider this dependence because it does not affect the [O/Fe]
features of the stars with [Fe/H] $> -3$ \citep{Goswami2000}. On
the \citet{grevesse1998} abundance scale, the iron and oxygen
abundances for the SN II ejecta are $Z_{\rm O, SN II}=13.2 \; \rm Z_{\rm
O, \odot}$ and $Z_{\rm Fe, SN II}=5.7 \; \rm Z_{\rm Fe, \odot}$
respectively, while for the SN Ia $Z_{\rm O, SNIa}=14.2\; \rm Z_{\rm O,
\odot}$ and $Z_{\rm Fe, SNIa}=430 \; \rm Z_{\rm Fe, \odot}$.  The
explosion energy of each SN of both types is $E_{\rm SN}=10^{51}$ erg.
   
\section{The Numerical Method}   
   
We run the simulations with the 3D BOH (BOlogna Hydrodynamics) 
hydro-code. The 3D code uses an Eulerian, second-order upwind scheme, 
in which consistent advection \citep{norman80} is implemented to 
reduce numerical diffusion. The 3D version of the code has been 
already used in different astrophysical contexts with satisfactory 
results \citep{marcolini2003, marcolini2004}. 
  
In all the simulations we adopt 3D Cartesian coordinates. The total 
number of mesh points is $160^3$. The central volume is covered by 
$100^3$ mesh points uniformly separated with $\Delta x=\Delta y=\Delta 
z=13$ pc; beyond this volume the linear mesh size increases 
geometrically in all directions with a size ratio of 1.14 between 
adjacent zones. In this way all the stellar region (where the SNe 
explode) is covered by the uniform grid and the grid edges are at 
large distances ($\sim 7$ kpc) in order to avoid that possible 
spurious perturbations originating at the boundaries may affect the 
solution in the central region. In all simulations outflow boundary 
conditions are enforced on all boundary planes. 
  
Besides solving the usual hydrodynamical equations, we also include 
two tracer variables representative of the SN II and SN Ia ejecta 
which are passively advected. Such tracers allow us to compute the 
chemical enrichment of the ISM due to the SN explosions.  The ejecta 
(and the energy) of each SN explosion is initially distributed into a 
volume with a radius of two mesh points around the SN location.

\begin{figure*}  
\begin{center}  
\psfig{figure=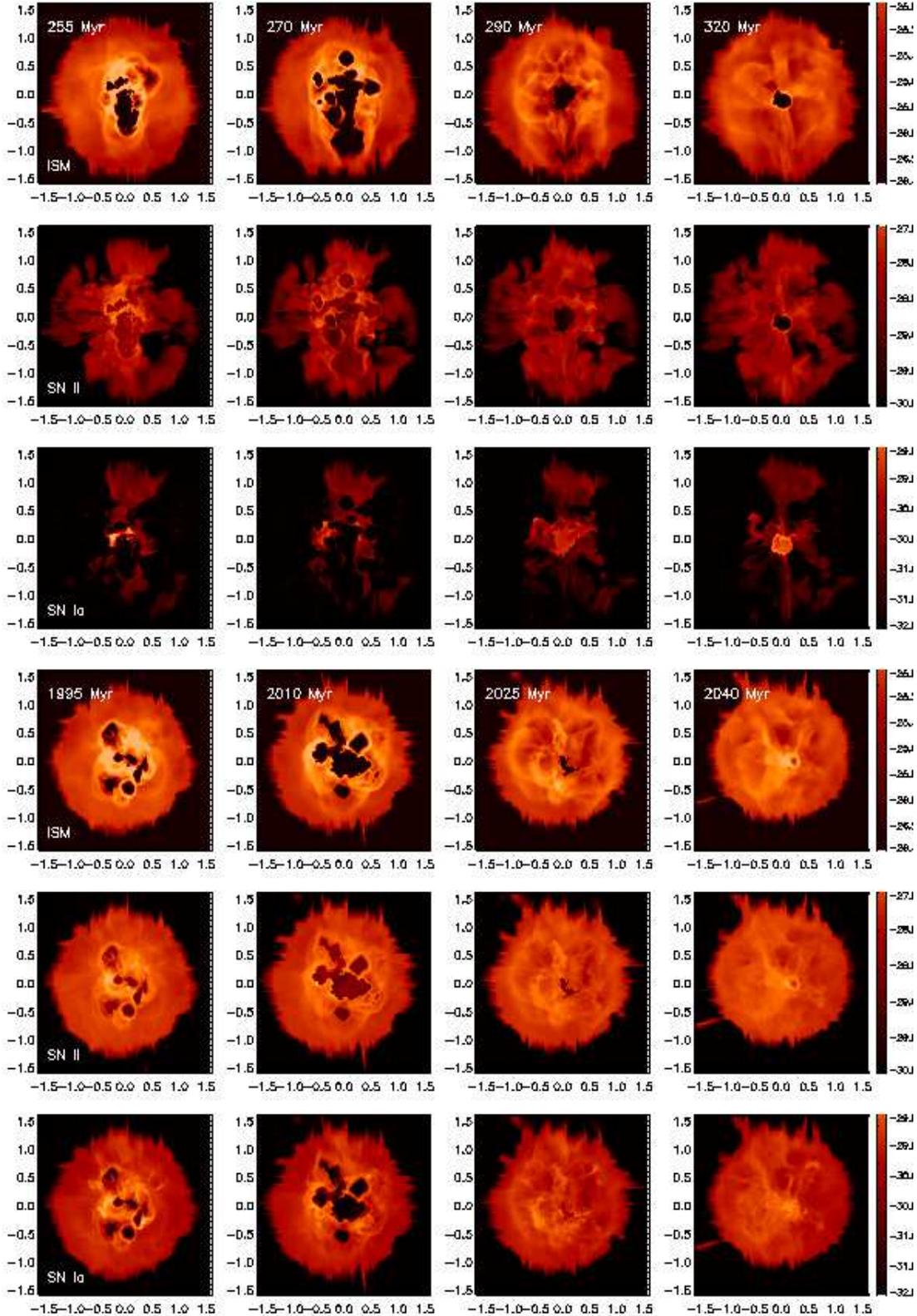}  
\end{center}  
\caption{Logarithm of the density distribution (g cm${^{-3}}$) of the  
ISM (first and fourth rows), SN II ejecta (second and fifth rows) and  
SN Ia ejecta (third and sixth rows) in the $z=0$ plane at different  
times for the reference model Draco-50. The first, second, third and  
fourth columns represent snapshots of the gas after a time interval  
$\Delta t=15$ Myr, 30 Myr, 45 Myr and 60 Myr from the occurrence of  
the latest instantaneous burst, respectively. Note that the time of  
the last column coincides with the occurrence of a new burst. Distances  
are given in kpc.}  
\label{fig:draco50}  
\end{figure*}

Radiative energy losses are taken into account considering the metal 
dependent cooling curve $\Lambda$ as calculated by 
\citet{sutherland1993}. In any case the temperature is never allowed 
to decrease below $10^4$ K.  In the Draco-S models, due to the low 
virial temperature of the ISM, the gas is allowed to cool down to 
$10^2$ K following, in this temperature range, the simple cooling 
curve given by \citet{rosenberg1973}. 
  
\section{The reference model: Draco-50}   
   
\subsection{Hydrodynamical evolution}

It can be shown (see appendix A) that, given the SNe II rate for unit 
volume of our model, the supernova remnants (SNRs) overlap forming a 
network of tunnels filled by hot rarefied gas with large filling 
factors after nearly $t=12$ Myr.  Actually, our numerical model 
confirms that after this time a large fraction of the stellar volume 
is filled by the hot rarefied gas of the SNRs' interior while the 
dense SNR shells form dense cold filaments after colliding one with 
another. Despite the galactic gravity, the great majority of these 
filaments move outward pushed by the shock waves of the successive SN 
II explosions. In the merging cavities the gas has a mean density as 
low as $10^{-28}$ g cm$^{-3}$ and a temperature of few $10^8$ K, and 
any further star formation is inhibited in these regions. Once the SNe 
II stop to explode (after 30 Myr since the beginning) the global 
cavity collapses and the ISM goes back into the potential well; this 
happens nearly 20 Myr after the last SN explosion. Note that the 
initial binding energy of the gas $E_{\rm bind} \sim 8.3 \times 
10^{52}$ erg is lower than the total energy $1.12 \times 10^{53}$ erg 
released by the SNe II (see Table~\ref{tab:sne}) after a single 
burst. The simulation thus shows that the radiative losses are 
substantial and prevent the evacuation of the gas, as shown in 
Fig.~\ref{fig:ebild} (upper panel). In this figure the binding, 
thermal and kinetic energies of the gas are reported. It is apparent 
that these latter energies are much lower than the binding energy 
although the energy released by SNe II after only a $single$ burst is 
$1.35 E_{\rm bind}$.  It is interesting to note that, despite its 
oscillatory behaviour, the mean amount of thermal energy of the gas 
remains constant, and does not increase in time as usually assumed in 
chemical evolutionary models. The consequences of this will be 
discussed further in Sections~\ref{sec:comparison} and 
\ref{sec:discussion}. 
 
\begin{figure}   
\begin{center}   
\psfig{figure=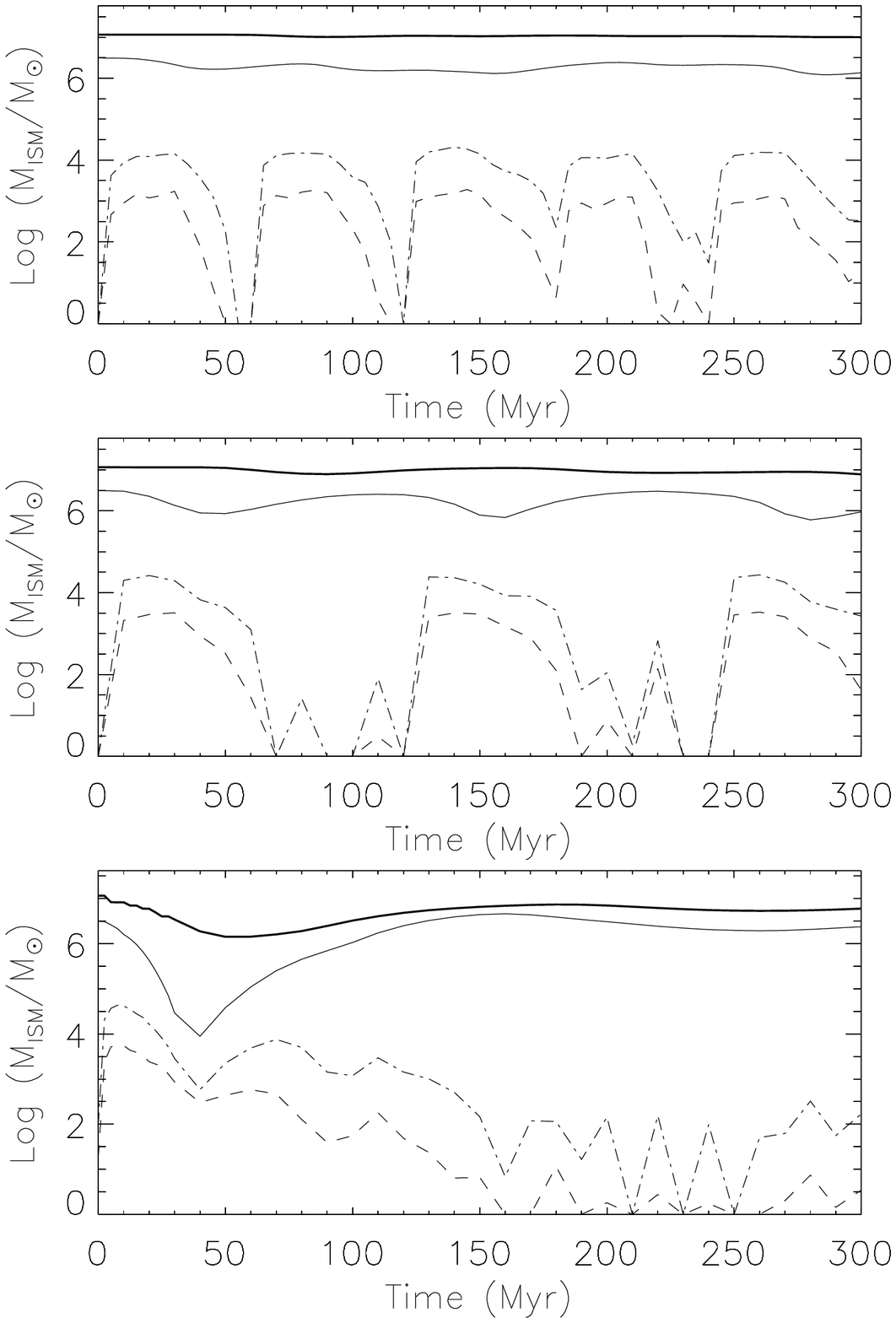,width=0.45\textwidth}   
\end{center}   
\caption{Time evolution of the cold ($T < 2 \times 10^4$ K) mass  
content inside the galactic (thick solid line) and stellar (thin solid  
line) regions. The dashed and dot-dashed lines refer to the warm ($2  
\times 10^4$ K $\le T \le 10^6$ K) and hot ($T > 10^6$ K) gas in the  
stellar region, respectively. The upper, middle and lower panels refer  
to the reference model Draco-50, Draco-25 and Draco-10, respectively.  
For the sake of simplicity only the evolution during the initial 300  
Myr is reported.}  
\label{fig:mhot}   
\end{figure}

\begin{figure}   
\begin{center}   
\psfig{figure=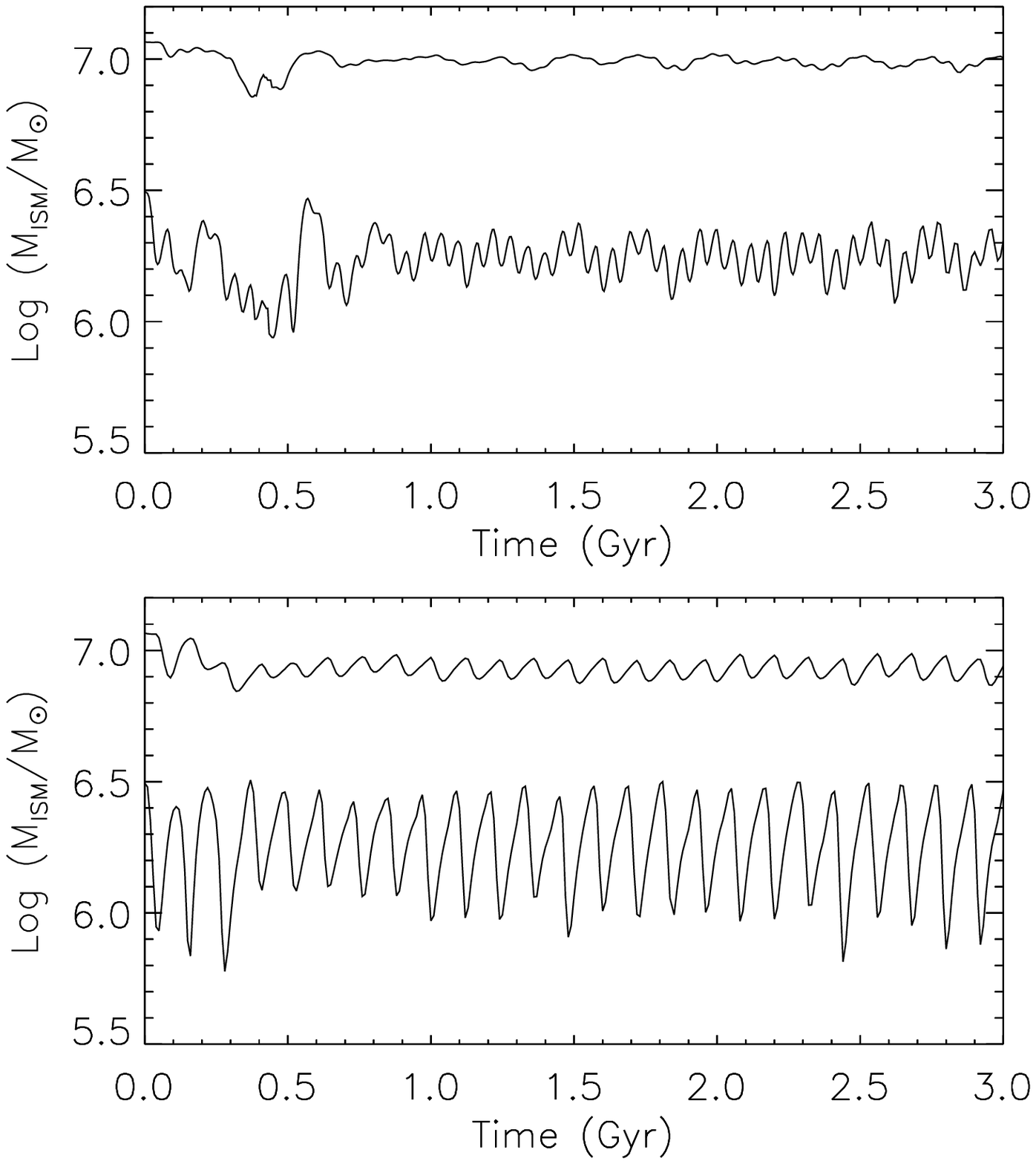,width=0.4\textwidth}  
\end{center}   
\caption{Time evolution of the mass content of the cold ISM ($T < 2  
\times 10^4$ K) inside the stellar (lower line) and galactic (upper  
line) regions for the reference model Draco-50 (upper panel) and model  
Draco-25 (lower panel).}  
\label{fig:mcold}  
\end{figure}

After 60 Myr the ISM distribution approximately recovers the initial 
condition, although turbulences and inhomogeneities are now present 
(see Fig.~\ref{fig:draco50}, fourth column). A second burst of star 
formation then occurs leading to a second sequence of SNII 
explosions. The gas undergoes a new cycle of merging bubbles which 
eventually collapse again.  The influence of the SNIa explosions on 
the general hydrodynamical behaviour of the ISM is not very important 
because, as apparent in Fig.~\ref{fig:snirate}, during a cycle of SN 
II explosions no more than 3-4 SNe Ia occur, only $\sim 3\%$ of the SN 
II number. Despite their little importance from a dynamical point of 
view, the role of SNe Ia is very relevant for the chemical evolution 
of the stars (see next subsection). 
  
Figure~\ref{fig:draco50} summarizes the evolution of the ISM and SN 
ejecta densities on the $z=0$ plane at several times.  In the first 
and fourth rows the formation, expansion, and successive interaction 
and re-collapse of the bubbles created by SNe II are illustrated. The 
central hole visible in the last panel of the first row is due to the 
explosion of a SN Ia occurred at $t \sim 297$ Myr.  Analogously, the 
second and fifth rows show the evolution of the SN II ejecta density, 
while the third and sixth rows show the evolution of the SN Ia ejecta 
at the same times. From this picture it is apparent that the gas may 
leave the central region of the galaxy but it is never lost. At an 
early stage the ejecta is distributed quite clumpy, but becomes more 
and more homogeneous with time as more SNe II explode and the 
turbulence diffuse it through the ISM. Note that the SN Ia ejecta 
appears to be distributed less homogeneously; the reason for this is 
the low SNe Ia rate. A deeper discussion on this point, and its 
consequences on the chemical evolution, will be presented in the next 
subsection. 
  
The periodic evolution of the ISM can be better seen in 
Fig.~\ref{fig:mhot} (upper panel) where the behaviour of the gas 
content of the galaxy is plotted as a function of the time. In this 
figure three different phases of the gas are shown: cold gas with $T < 
2 \times 10^4$ K, warm gas with $2 \times 10^4$ K $\le T \le 10^6$ K and 
hot gas with $T > 10^6$ K. This distinction helps us to understand the 
thermal evolution of the gas and the exchange among the cold, warm and 
hot phases. The warm and hot gas phases reside mainly in the stellar 
region, and their time evolution is indicated by the dashed and 
dot-dashed lines, respectively. 
  
For the sake of simplicity, in Fig.~\ref{fig:mhot} we show only the 
evolution during five cycles in the time interval $0 < t < 300$ 
Myr. The evolution over the entire period of 3 Gyr shows essentially 
the same periodicity, as can be seen in Fig.~\ref{fig:mcold} (upper 
panel) where the cold gas mass content of the galaxy is plotted up to 
3 Gyr. 
  
Despite its oscillatory behaviour, nearly 30\% of the initial gas is 
located beyond the galactic region. However, this gas is not lost by 
the galaxy but rather expands at larger distances remaining 
gravitationally bounded. Such a gas is obviously the most susceptible 
to be lost by ram pressure and tidal interaction with the Milky Way 
\citep[see e.g.][]{marcolini2003, sofue1994, mastropietro2005, 
mayer2005}. Indeed, since SNe feedback is not able to evacuate the ISM 
from the galaxy, in this scenario we must invoke an external 
mechanism, such as ram pressure stripping and/or tidal interaction, to 
remove the gas and thus terminate the star formation. Ram pressure due 
to gaseous haloes of the Milky Way and M31 may also explain the 
observed correlation between stellar content and the galactocentric 
distance of dwarf galaxies \citep{vandenbergh1994}.

\begin{figure*}   
\begin{center}   
\psfig{figure=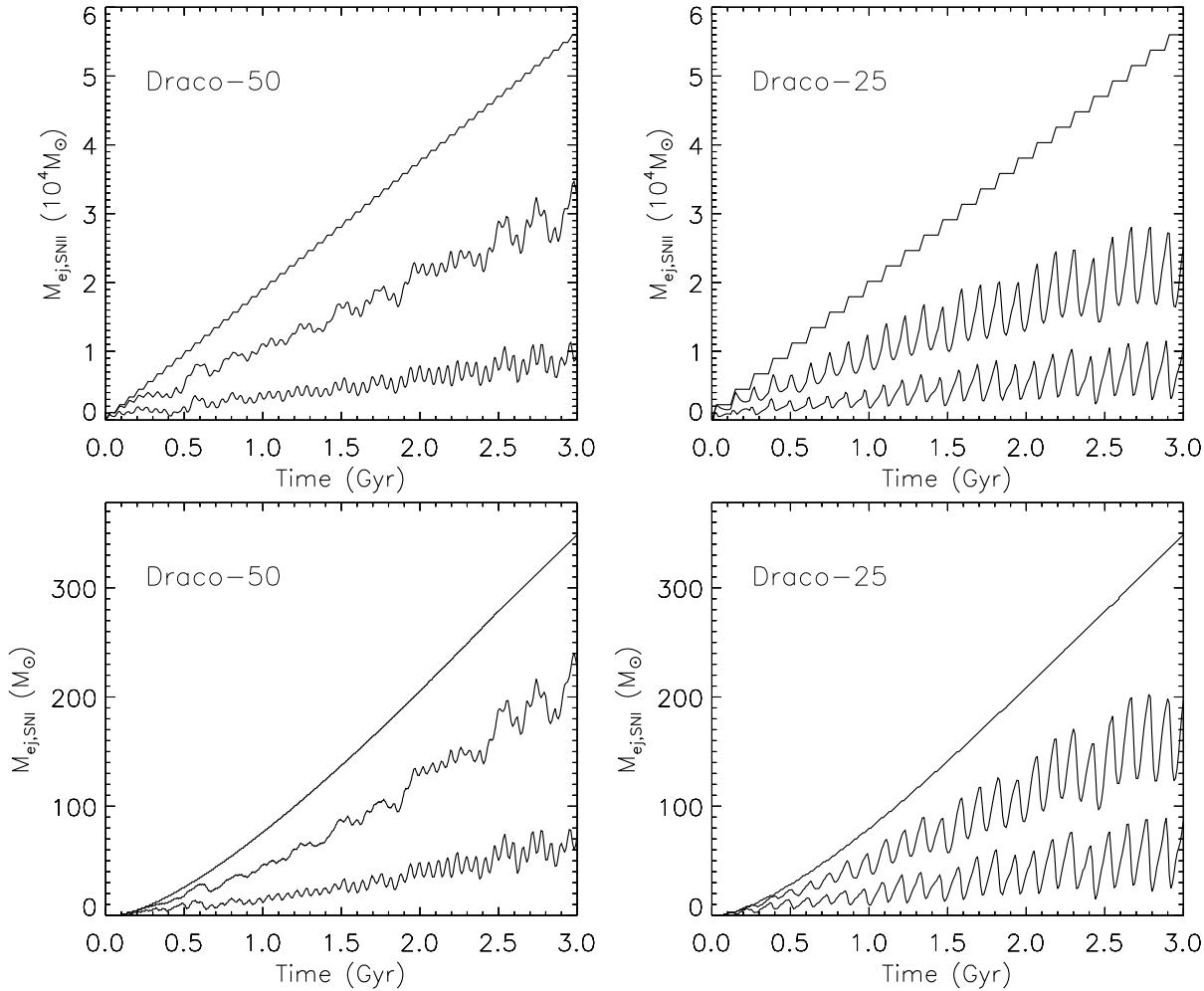}
\end{center}   
\caption{Time evolution of the mass of the SN II (upper panels) and SN 
Ia (lower panels) ejecta inside the stellar (lower lines) and galactic 
(middle lines) regions. The upper stair shaped lines in all panels 
represent the total amount of SN ejecta expelled during 3 Gyr.  The 
first and the second columns refer to the models Draco-50 and 
Draco-25, respectively.} 
\label{fig:tracer}   
\end{figure*}

\subsection{Chemical evolution}  
\label{sec:chemical_evolution} 
 
The upper line in the top left panel of Fig.~\ref{fig:tracer} 
represents the total amount of SN II ejecta expelled during 3 Gyr; the 
other two lines illustrate the evolution of the ejecta content inside 
the galactic (middle line) and the stellar (lower line) regions. The 
SN Ia ejecta has a similar behaviour (lower left panel). Note that 
along the entire evolution the fraction of the ejecta present inside 
the stellar region remains very low ($\sim 18\%$ after 3 Gyr).  This 
is the amount of metals which contributes to the metallicity of the 
forming stars. A large fraction of the ejecta is pushed at larger 
distances by the continuous action of the SN explosions. Indeed, 
nearly half of the ejecta still resides inside the galactic volume 
where it mixes with the ISM. The missing $\sim$ 40\% of the ejecta is 
still bounded and is mostly found in the neighbourhood of the galaxy, 
although a part of it is lost through breaches in the cold gas.

\begin{figure*}   
\begin{center}   
\psfig{figure=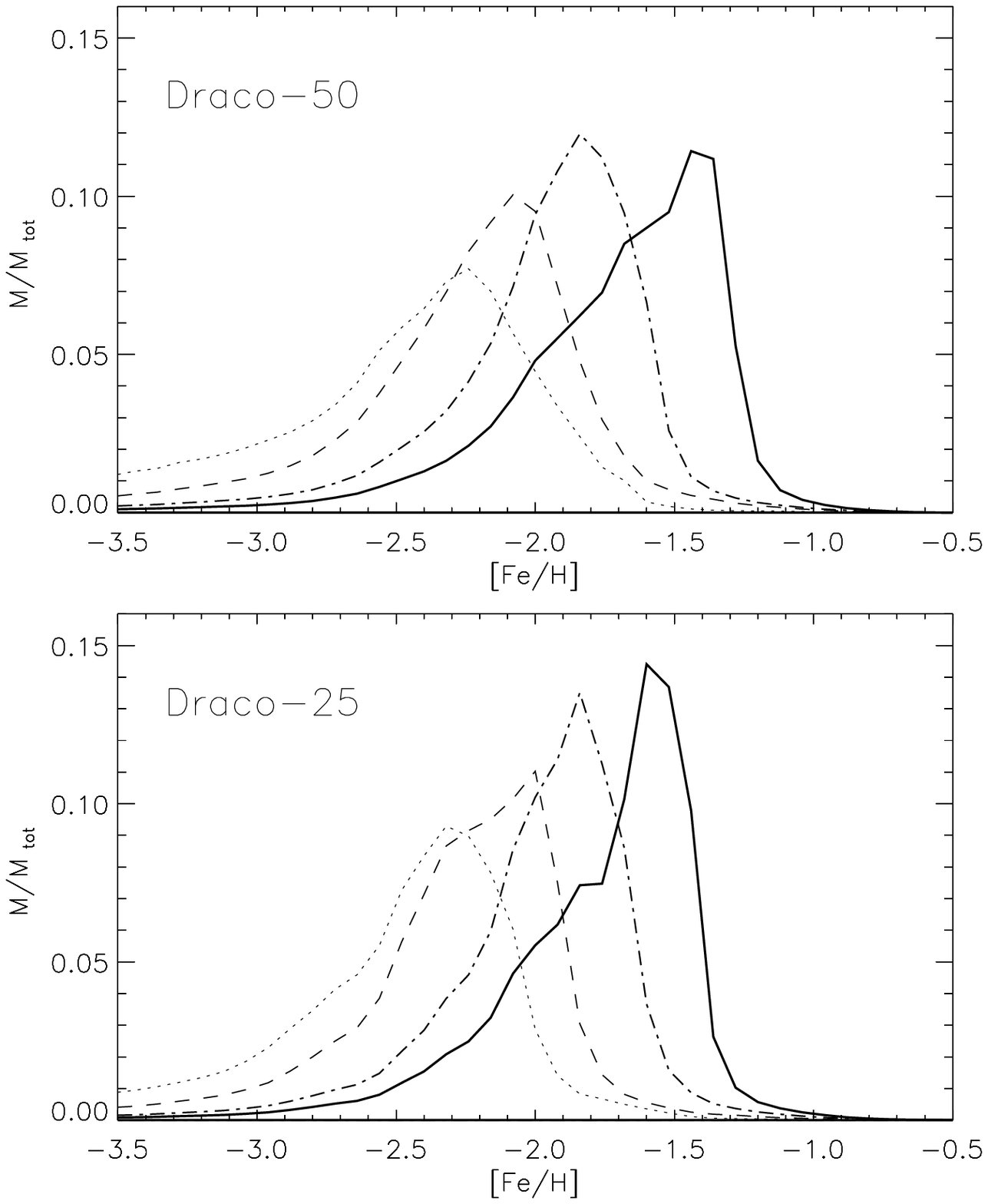}   
\end{center} 
\caption{[Fe/H] distribution function of the long-lived stars 
formed up to four different times for the reference model Draco-50 
(upper panel) and model Draco-25 (lower panel). Dotted line: 240 Myr; 
dashed line: 600 Myr; dot dashed line: 1.5 Gyr; solid line: 3 Gyr.} 
\label{fig:zstelle} 
\end{figure*} 
 
Further insight on the chemical evolution is given by 
Fig.~\ref{fig:zstelle}. In the upper panel we plot the [Fe/H] 
distribution function (MDF) of the long-lived stars (with a mass $\le 
0.9$ M$_{\odot}$), i.e. the mass fraction of these stars as a function 
of their [Fe/H]. The distribution is given at four different times  
in order to show its time evolution. 
 
We construct the MDF as follows. At every starburst the stellar mass 
in each ith computational cell of volume $V_{\rm i}$ is increased by 
an amount $\Delta M_{\star,\rm i}=\Delta \rho_{\star, \rm i} V_{\rm 
i}$, where $\Delta \rho_{\star, \rm i} =\rho_{\star, \rm i}/N_{\rm 
burst}$ represents the density of the newly formed stars 
(c.f. Section~\ref{sec:supernovae_explosions}).  Stars are not created 
in those cells where $T > 2 \times 10^4$ K.  The number of such cells 
in each starburst is however small and the general total stellar 
profile is still well described by equation~\ref{equ:star_density}. 
The newly formed stars are assumed to have the same chemical 
composition of the ISM in the ith cell at the moment of the burst: 
 
\begin{equation}   
Z_{\star, \beta}= (\rho_{\rm ISM} Z_{\rm ISM,\beta}^{0} +   
\rho_{\rm SNII} Z_{\rm SNII,\beta} +   
\rho_{\rm SNIa} Z_{\rm SNIa,\beta})/\rho_{\rm tot},  
\end{equation}   
\noindent  
where $\rho_{\rm tot}=\rho_{\rm ISM}+\rho_{\rm SNII}+\rho_{\rm SNIa}$,  
and $Z_{\rm ISM,\beta}^{0}$ is the initial ISM abundance of the  
generic element $\beta$.   
  
At early times the MDF is rather broad reflecting an inhomogeneous 
distribution of the SN ejecta inside the galaxy because of the still 
low number of SNe explosions. As the number of the bursts increases, 
the ejecta becomes distributed more and more homogeneously, and the 
MDF of the newly formed stars peaks around the mean value 
$\langle$[Fe/H]$\rangle$, which increases with time. For instance, the 
stars forming in the first burst are distributed rather shallowly in 
the range $ -4 < $[Fe/H]$ < -1.9$, while those formed in the last 
burst are clustered in the range $-1.5 < $[Fe/H]$ < -1.2$. 
  
\begin{figure}   
\begin{center}   
\psfig{figure=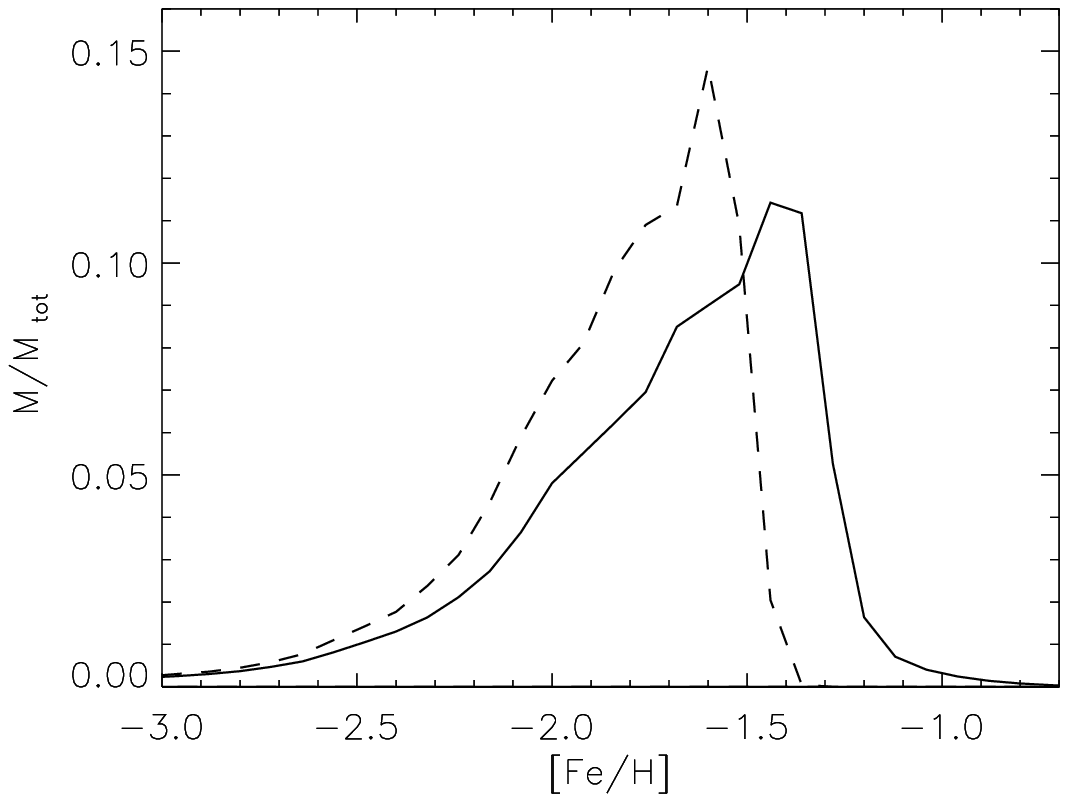,width=0.45\textwidth}   
\psfig{figure=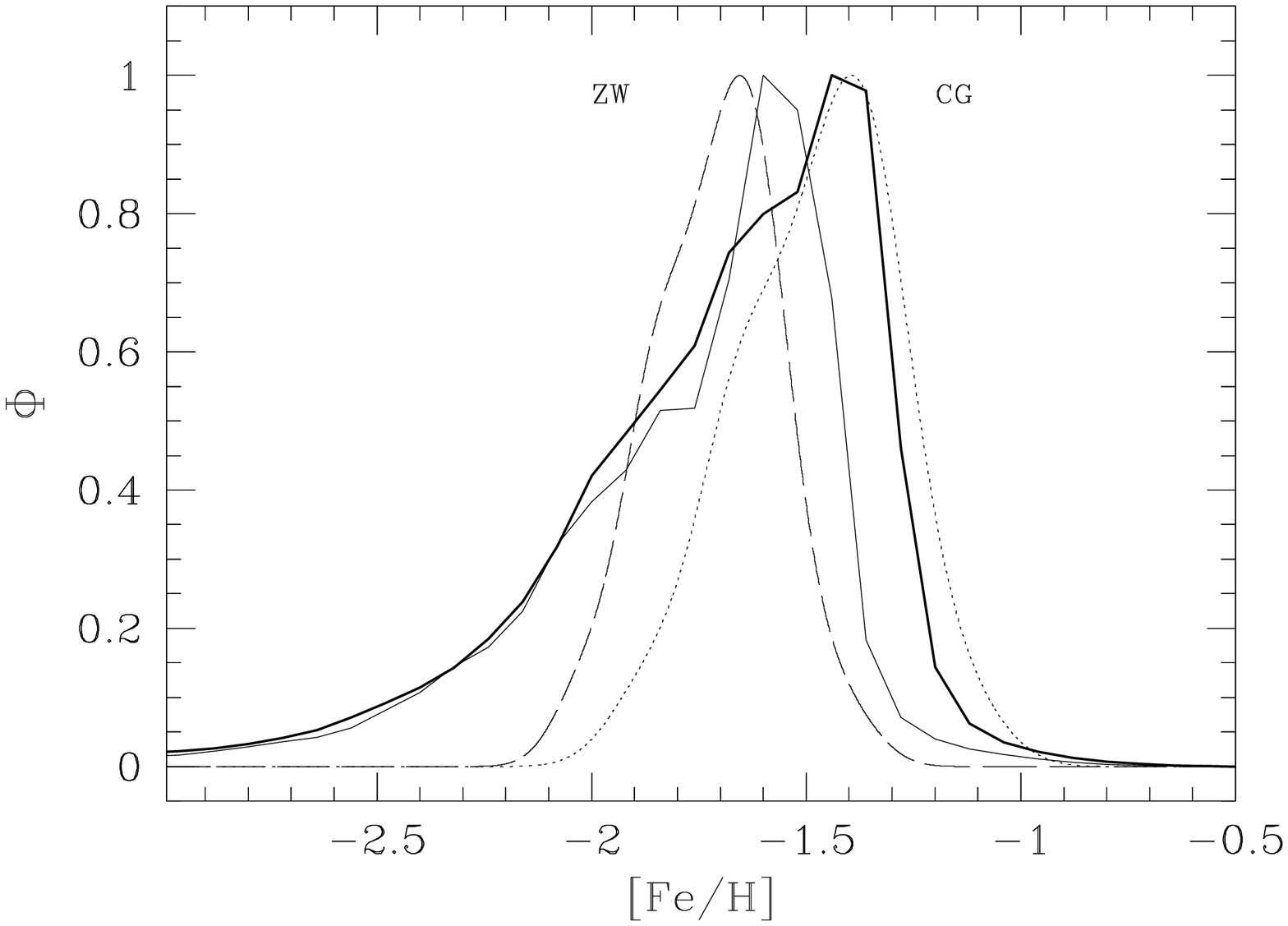,width=0.45\textwidth}   
\end{center}   
\caption{Upper panel: [Fe/H] distribution function of the stars 
at the end of the simulation for the reference model Draco-50 (solid 
line). The dashed line illustrates the contribution due to SNe II 
only. Lower panel: normalized MDF at t=3 Gyr for the models Draco-50 
(thick solid line) and Draco-25 (thin solid line) compared to the 
observative MDFs of Draco \citep{bellazzini2002} in the Zinn-West 
(dashed line) and Carretta-Gratton (dotted line) metallicity scales.} 
\label{fig:zstellenosni}   
\end{figure}

In Fig.~\ref{fig:zstellenosni} (upper panel) we show the final MDF 
together with the contribution due to SNe II only. Confronting these 
two types of contributions, we note that the effect of the enrichment 
due to SNe Ia produces a larger dispersion particularly evident at the 
high [Fe/H] tail; this spread lowers the MDF maximum because the curve
integrals (i.e. the total mass of the stars) must be the same.   

The reason for the spread is given by the highly 
inhomogeneous distribution of the SN Ia ejecta.
Although the number of SNe Ia explosions after 3 Gyr is only 4.5\% of 
that of SNe II, each of them expels 10 times more iron mass, and thus 
their contribution to the ISM metallicity is relevant. This 
contribution is particularly conspicuous in the high [Fe/H] tail 
of the MDF because, contrary to SNe II, SNe Ia may explode at any 
time, and in particular during the quiescent phases of SNe II.  
During these periods the ISM re-collapses and the SN Ia ejecta expelled in 
these phases remains more localized because of the larger density of 
the ambient medium. Actually, a rough calculation indicates that the 
SN Ia porosity (see appendix A) at this stage is $Q \sim 0.04$ 
(assuming a mean $Z=0.01 \; \rm Z_{\odot}$). This means that the SNe Ia 
remnants are located quite apart one from another. Stars forming in 
these regions enriched by SNe Ia have an enhanced [Fe/H] ratio 
compared to the stars forming in the rest of the stellar volume, and 
are responsible of the high [Fe/H] tail of the MDF. 
 
At the end of the simulation we obtain a mean value 
of $\langle$[Fe/H]$\rangle$=-1.65 with a spread of $\sim$ 1.5 dex, in 
reasonable agreement with observations \citep{lehnert1992, 
shetrone2001, aparicio2001, bellazzini2002}, while the 
maximum value of the distribution occurs at [Fe/H]$\sim$-1.5. 
 
In the lower panel of Fig.~\ref{fig:zstellenosni} the final MDF is 
compared with the MDF obtained by \citet{bellazzini2002}, using both 
the abundance scale by \citet{carretta1997} and \citet{zinn1984}. The 
calculated MDF is in good agreement with the observed one at the high 
[Fe/H] end. The tail of our MDF in the low [Fe/H] range is instead higher  
and less steep than the observed MDF. This is likely due to our  
simple SFH (bursts of equal strength) assumed.  A more realistic SFH,  
with weaker early bursts, could depress our final MDF at low  
[Fe/H]. Also, the relative insensitivity of the method adopted by  
\citet{bellazzini2002} to the most metal poor stars, possibly reduces  
the discrepancy indicated in Fig.~\ref{fig:zstellenosni}.  
  
\begin{figure*}  
\begin{center}  
\psfig{figure=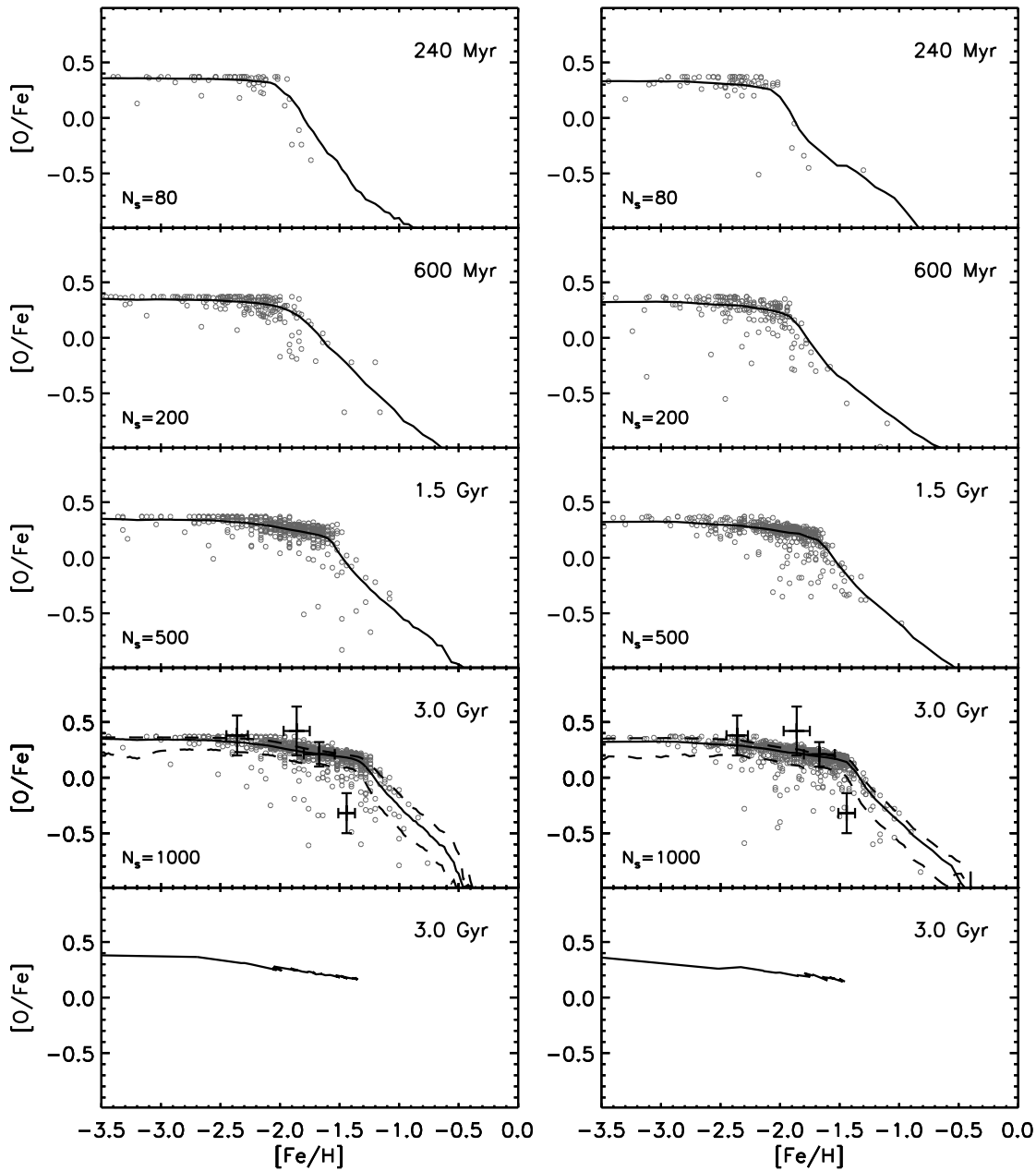}
\end{center} 
\caption{Abundance ratio [O/Fe] plotted against [Fe/H] of $N_{\rm s}$ 
sampled stars at four different times for the reference model Draco-50 
(left column) and model Draco-25 (right column). At each time $N_{\rm 
s}$ represents a fraction $10^{-4}$ of the total number of stars at 
that time. The solid line represents the mean value of the [O/Fe] 
distribution for any fixed [Fe/H]. The dotted lines in the fourth 
panels represent the 1 $\sigma$ spread of this distribution. In the 
fourth panels we also show the values obtained by 
\citet{shetrone2001}.  In the lowest panels the same diagrams are 
obtained mimicking a one-zone model, i.e. assuming that the stars 
forming at each time have all the same metallicity, given by the 
spatial mean metallicity of the ISM at that time.} 
\label{fig:otofe} 
\end{figure*}

ThE effect of the MDF tail at high [Fe/H] discussed above 
is particularly evident in 
Fig.~\ref{fig:otofe} where the diagram [O/Fe]-[Fe/H] is shown at four 
different times. The open circles form a statistically representative 
sample of the stellar distributions in [O/Fe] and [Fe/H]. At each time 
the sample is composed by a fraction $10^{-4}$ of the total number of 
stars. The solid line represents the mean value of the [O/Fe] 
distribution for any fixed [Fe/H].  The dotted lines in the fourth 
panel indicate the 1 $\sigma$ spread of this distribution.  At any 
time a plateau at low values of [Fe/H] is delineated by the solid 
line, followed by a sharp decrease. With time the break in the 
distribution moves toward higher values of [Fe/H]. The plateau is due 
to the initially dominant contribution of SNe II to the Fe abundance: 
being the oxygen formed mainly by SNe II, the ratio [O/Fe] must remain 
nearly constant as the iron increases. The small negative gradient of 
the plateau is due to the slowly growing contribution in the Fe 
enrichment by SNe Ia. The sharply decreasing branch at higher [Fe/H] 
is due to stars formed in the regions of ISM recently polluted (mostly 
by iron) by SNe Ia. For this reason the stars populating the solid 
line in the decreasing branch are distributed diagonally nearly 
following a -1 slope.  A glance at Fig.~\ref{fig:zstelle} and 
Fig.~\ref{fig:otofe} shows that the stars on the decreasing branch 
populate the MDF high [Fe/H] tail, while the majority of the 
stars occupies the high [Fe/H] edge of the plateau in the 
[O/Fe]-[Fe/H] diagram. 
  
In the fourth panel of Fig.~\ref{fig:otofe} data from 
\citet{shetrone2001} are reported. The four stars seem more uniformly 
distributed in [Fe/H] than our sample. This is because these four 
stars have been selected in order to span the largest range in [Fe/H] 
(Shetrone, private communication); thus their distribution has no 
statistical meaning and is not expected to follow our sample 
distribution. We note that the point with the lowest [O/Fe] is 
displaced by $\sim 0.5$ dex to the left relative to the solid line 
denoting the descending branch. This discrepancy does not represent a 
severe drawback for our model; there are in fact several explanations 
for it. On the theoretical side, had we started with a more extended 
gaseous halo, we would have found a lower mean value of [Fe/H], moving 
our sample to the left in the [O/Fe]-[Fe/H] diagram. On the 
observative side, the derived mean [Fe/H] value depends on the 
metallicity scale adopted (cf. the lower panel of 
Fig.~\ref{fig:zstellenosni}): for Draco it ranges in the interval 
$-2<\langle $[Fe/H]$ \rangle < -1.6$ \citep{aparicio2001}. 
\citet{shetrone2001} (whose data reported in our figure are, to our 
knowledge, the only published for Draco) find $\langle$[Fe/H]$\rangle= 
-2.0$: it is reasonable to hypothesize that, with a different choice 
of the metallicity scale, the observed stars reported in 
Fig.~\ref{fig:otofe} would shift to the right by an amount very nicely 
fitted by our model. 
 
\begin{figure*}  
\begin{center}   
\psfig{figure=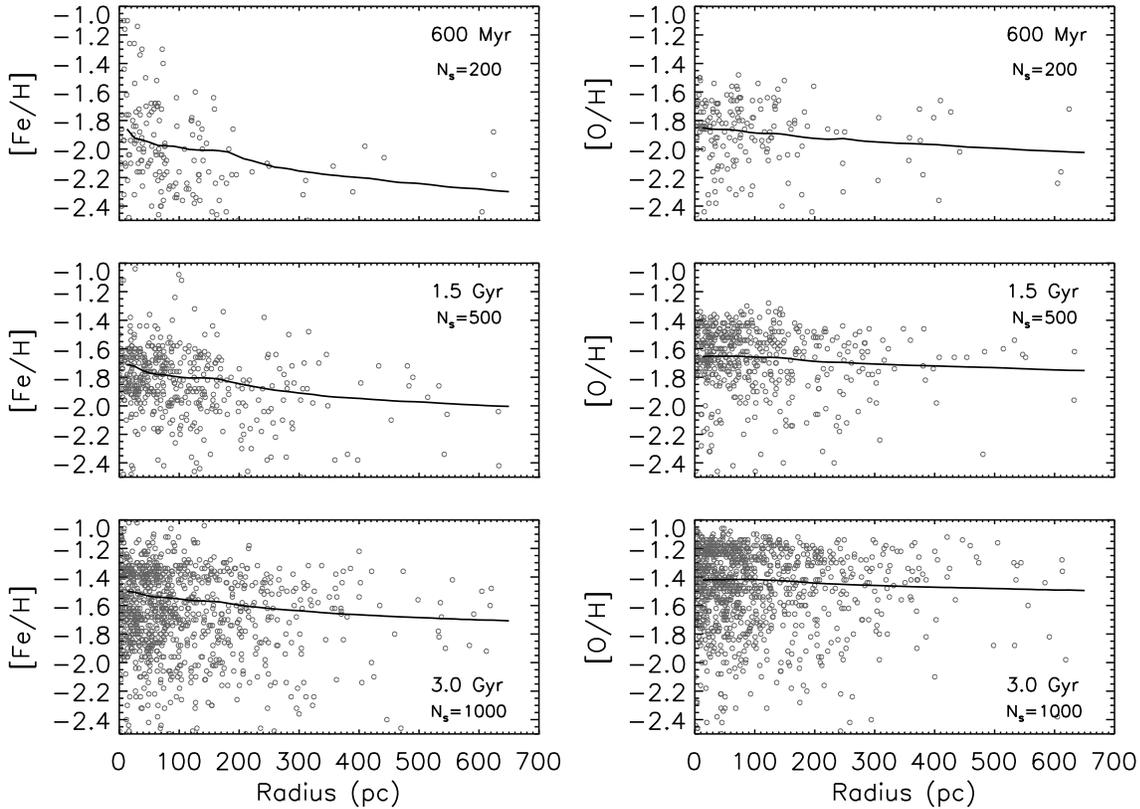,width=0.9\textwidth}   
\end{center}   
\caption{Radial distribution of [Fe/H] (left panels) and [O/H] (right 
panels) at three different times for the reference model Draco-50. 
The solid lines represent the radial profile of the mean values.  
$N_{\rm s}$ has the same meaning as in Fig.~\ref{fig:otofe}.}  
\label{fig:zraggio}   
\end{figure*}

It is interesting to analyse the radial distribution of the ratio 
[Fe/H] within our model. Age/metallicity gradients have been claimed 
by some authors \citep{harbeck2001, lee2003, bellazzini2005} to be 
present in several dSphs while others do not find them 
\citep{hurley1999, aparicio2001, carrera2002, babusiaux2005}.  The 
radial dependence of the ratios [Fe/H] and [O/H] at three different 
times are shown in Fig. \ref{fig:zraggio} where the circles form a 
statistically representative sample of the stars and the solid lines 
represent the radial profiles of the mean values of [Fe/H] and 
[O/H]. A big metallicity spread of $\sim 1.5$ dex is present at any 
radius \citep[consistent with some observations, e.g.][]{tolstoy2004}, 
and the profiles have been obtained splitting the galaxy in 50 
spherical shells and averaging inside each shell. 
 
A slight metallicity gradient is actually present and reduces with 
time. After 3 Gyr the difference between the central and the outer 
region is 0.2 dex in [Fe/H], consistent with observations 
\citep{harbeck2001}, while the [O/H] gradient is virtually absent. 
The origin and the evolution of the gradients are given by the 
following reasons. The ejecta are subject to two opposite effects: on 
one hand they are released from SNe distributed following a King 
profile, thus more concentrated toward the centre, and on the other 
hand the hydrodynamics tend to distribute them more uniformly. With 
time the mean gas metallicity increases while becoming more and more 
uniformly distributed in space, and the metallicity gradient, 
essentially given by the freshly expelled metals, becomes less and 
less important. 
 
We point out that the [O/H] gradient is shallower than that of 
[Fe/H]. In fact, a substantial amount of iron present in the ISM is 
produced by SNe Ia whose ejecta, as discussed above, tends to remain 
more localized toward the centre (in particular during the quiescent 
phases) than the oxygen, which is expelled mainly by SNe II.

\section{Other models}  
  
\subsection{Draco-25}  
  
In order to understand the influence of the SFH on the chemical and 
dynamical evolution of the ISM, we computed a model similar to the 
reference one but with a time interval $\Delta t_{\rm burst}=120$ Myr 
between two successive bursts. As a consequence ${\cal R}_{\rm SN 
II}=7.47 \times 10^{-6}$ yr$^{-1}$, doubled respect to the rate of the 
reference model. 
  
\begin{figure*}   
\begin{center}   
\psfig{figure=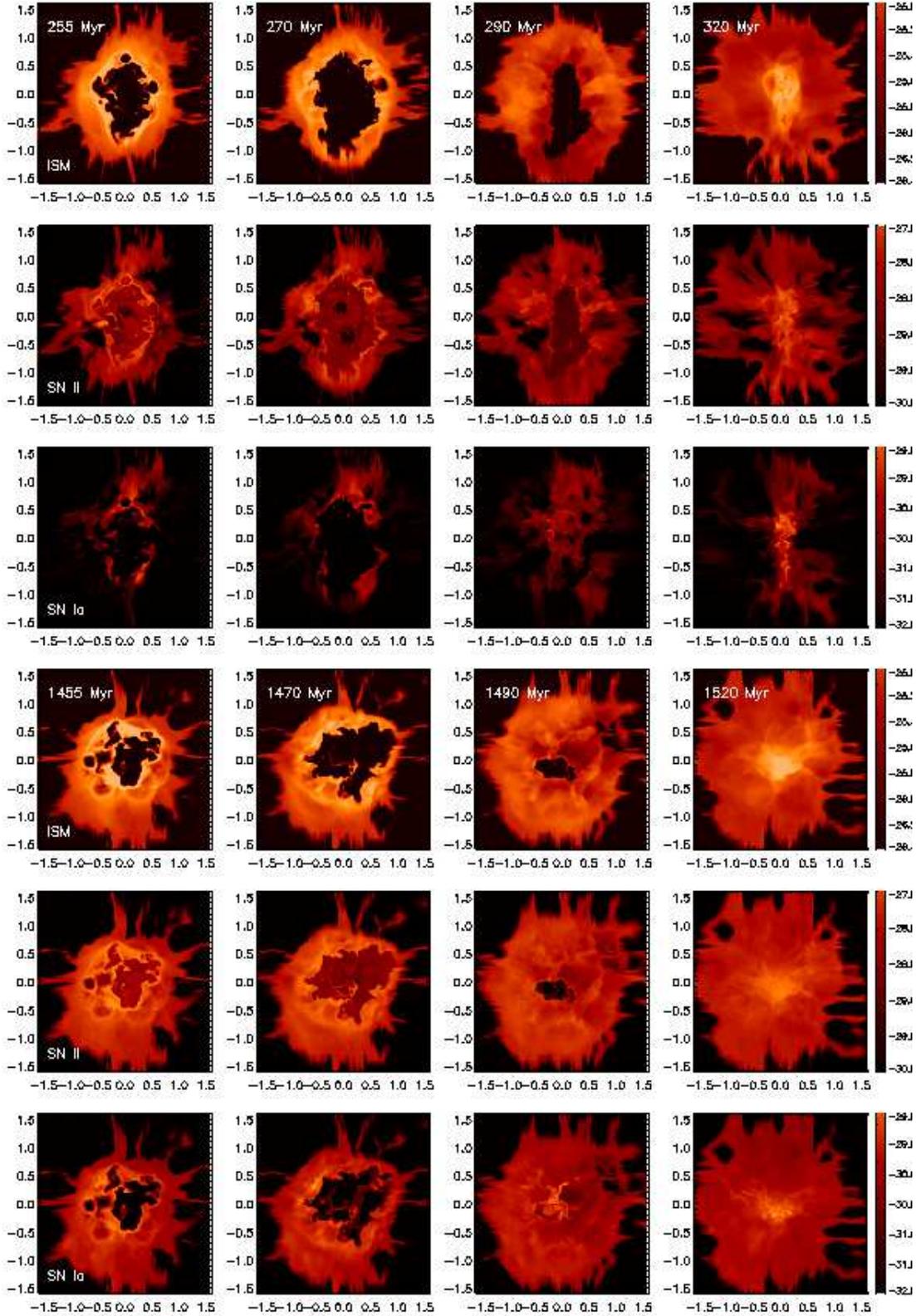}   
\end{center}  
\caption{Logarithm of the density distribution (g cm$^{-3}$) of the  
ISM (first and fourth rows), SN II ejecta (second and fifth rows) and  
SN Ia ejecta (third and sixth rows) in the $z=0$ plane at different  
times for the model Draco-25. The first, second, third and fourth  
columns represent snapshots of the gas after a time interval $\Delta  
t=15$ Myr, 30 Myr, 50 Myr and 80 Myr from the occurrence of the latest  
instantaneous burst. Distances are given in kpc.}  
\label{fig:draco25}   
\end{figure*}

Due to this higher rate, the size of the cavities forming in the ISM 
inside the stellar region are larger (see Fig.~\ref{fig:draco25}). 
The gas at the periphery of the galaxy results to be more extended and 
less dense than in the reference model (compare Fig.~\ref{fig:draco25} 
and Fig.~\ref{fig:draco50}). However, although the total SN energy 
released during one burst is 2.7 times the initial ISM binding energy, 
radiative losses are still effective enough to impede the gas to be 
lost by the galaxy.  Given the larger value of $\Delta t_{\rm burst}$, 
the ISM has more time to pile up in the central region during the 
re-collapse before the next burst, thus reaching higher densities than 
in the Draco-50 model. In this region, which has actually the 
dimension of the stellar region, conditions are recovered for a new 
burst of star formation. 
 
The evolution of the ISM mass content inside the galactic and stellar 
regions is shown in Fig.~\ref{fig:mcold} (lower panel). Contrary to 
the reference model, here the oscillations are much more regular 
because of the larger value of $\Delta t_{\rm burst}$. 
 
The fraction of the ejecta present inside the stellar region at the 
end of the simulation (see Fig~\ref{fig:tracer}) is slightly smaller 
than in the reference model. Indeed, being the single bursts more 
powerful, the hot metal rich gas has more chances 
to leave the galaxy. As a consequence, the MDF peaks at a lower value 
of [Fe/H]$\sim -1.6$.

The chemical enrichment of the galaxy is synthesized in the lower 
panels of Fig.~\ref{fig:zstelle} and Fig.~\ref{fig:zstellenosni}, and 
the right column of Fig~\ref{fig:otofe}. We obtain a final value 
$\langle$[Fe/H]$\rangle -1.7$ with a spread of $\sim$ 1.5 dex. The 
descending branch in the [O/Fe]-[Fe/H] diagram is now closer to the 
star with the lowest value of [O/Fe]. In general, however, these 
results show that the chemical evolution is slightly dependent on the 
adopted SFH. 
 
In Fig.~\ref{fig:zraggio25} we plot the final radial distribution of 
the ratio [Fe/H] and [O/H], as well as the radial profile of their 
mean values. As for the reference model, the chemical gradient results 
to be very shallow at the beginning of the simulation, and tends to 
disappear with time.  
 
\subsection{Draco-10} 
 
In this model only ten bursts occur during 3 Gyr. They are separated 
in time by 300 Myr, and each of them is five times more powerful than 
those of the reference model. 
 
Each single burst tends to expand the ISM more than in the previous 
models, but radiative losses are still able to impede the gas to leave 
the galaxy. Indeed, as apparent in Fig.~\ref{fig:mhot} (lower panel), 
the mass content of the cold gas inside the stellar region may become 
as low as $10^4$ M$_{\odot}$ just after the SNe II activity, but then 
grows again to the initial value during the SNe II quiescent phase. 
The gas continues to pile up toward the centre until, at $t \sim 150$ 
Myr, rebounds and its content decreases slightly (see the depression 
of the temporal mass profile at $t=250$ Myr). 
  
After 600 Myr the chemical properties of the forming stars differ only 
slightly from those of the previous models at the same time, 
indicating that the chemical evolution is rather insensitive to the 
chosen SFH. For this reason, given the huge amount of CPU time 
required to complete the model, we stopped our simulation after 
$t=600$ Myr. 
  
\subsection{Draco-S}  
 \label{sec:draco-s} 
  
As discussed in Section~\ref{sec:model}, these models have been run to
test the possibility that the dark halo could be lighter than the one
we assumed in the reference model. In fact, relaxing the hypotheses of
isotropic velocity dispersions and virial equilibrium, one obtains
mass-to-light ratios much lower than commonly adopted \citep[see
e.g.][for the case of UMi]{gomez2003}. The non baryonic mass content
(and consequently the baryonic one) in the model Draco-S is assumed to
be only one third of the standard model (c.f. Table 1 and 2): $M_{\rm h}=
2.2 \times 10^7$ M$_{\odot}$ \citep{mateo98} which gives a $total$
$M/L_{\rm V}=80$ M$_{\odot}/\rm L_{\odot}$.

We explored the same SFH already discussed for the previous models
with $N_{\rm burst}=10$, 25 and 50. In Fig.~\ref{fig:massa80} the
evolution of the mass content inside the dark matter halo is shown for
all the three models. The galaxy gets rid of its ISM in all models on
a timescale which increases with the number of bursts.  This is due to
the fact that when only few bursts occur, each of them is rather
powerful (c.f. section 2.2), and can eject the ISM sooner than many
weaker bursts. In all cases the gas is lost before $t \sim 250$ Myr, a
time too short to be compatible with the star formation duration
inferred by the observations \citep[e.g.][]{mateo98, dolphin2002}.
Our results (in the limit of their assumptions, e.g. the ISM mass
content and distribution are related to the DM ones and self gravity
is neglected) indicate that the total $M/L$ must be quite high and the
dark halo must extend far beyond the stellar component.


\section{Comparison to other models}  
 \label{sec:comparison}

The SFH and the chemical enrichment of Draco and other dSphs have been 
discussed in terms of purely one zone chemical or cosmological models 
by several authors.

\begin{itemize} 
 
\item In their one zone model LM find that the plateau in the 
  [O/Fe]-[Fe/H] plane is representative of SNe II enrichment, although 
  their plateau level is higher than that found in our models because 
  LM are able to distinguish the contribution to the metal production 
  by SNe II of different masses, while we assume that every SN II has 
  the same (average) chemical composition. The sudden [O/Fe] decrease 
  at larger values of [Fe/H] is interpreted by LM as a consequence of 
  a galactic wind which decreases the amount of available gas; the 
  formation of new stars is thus slowed down and so is the production 
  and the injection of $\alpha$ elements into the ISM by SNe II. SNe 
  Ia, instead, have progenitors with much longer evolutionary 
  timescales \citep{matteucci1986a}, and continue to inject iron 
  lowering the [O/Fe] ratio.  On the contrary, in our models we do not 
  have the formation of any wind and the change of the slope in the 
  [O/Fe] vs. [Fe/H] diagram is given, as discussed in 
  Section~\ref{sec:chemical_evolution}, by the inhomogeneous 
  distribution of the SNIa ejecta.  In both models the descending 
  branch is less populated than the plateau. In the LM models this 
  happens because the most metallic stars form during the wind phase, 
  when the star formation rate is greatly reduced; more metallic stars 
  can be obtained increasing the star formation efficiency, although 
  this leads to an overestimate of the [O/Fe] ratio. 
 
It is interesting to discuss further the differences between our 
models and those of LM. In these latter models the star formation rate 
depends on the ISM density and varies with time, while the SN II 
efficiency in heating the ISM has a $fixed$ value assumed ``a priori'' 
($\epsilon_{\rm SN II}=0.03$ in their notation), i.e. only a small 
fraction $\epsilon_{\rm SN II}$ of the energy of the SNe II explosions 
is transferred to the surrounding gas, the rest being radiated 
away. This energy fraction is constantly accumulated in the ISM, and 
the occurrence of a galactic wind at a certain time is 
unavoidable. The choice of the value of $\epsilon_{\rm SN II}$ is 
crucial for the behaviour of the models. An increase/decrease in its 
value anticipates/retards the occurrence of the wind (for a fixed 
potential well). Thus, an appropriate choice of the ratio between 
thermalized SN II energy and potential well depth is required to 
obtain a reasonable duration of the star formation. Our models, in a 
sense, work opposite to those of LM. Intensity and duration of the 
star formation are $fixed$ ``a priori'', while the fraction of SNe II 
energy transmitted into the ISM is regulated by hydrodynamics and 
radiative losses. In our models winds either occur very soon or do not 
occur at all. In this scenario, a prolonged star formation period as 
that inferred for Draco (and other dSphs) suggests that the 
termination of star formation is due to an external cause such as gas 
stripping by the Galaxy \citep[e.g.][]{mayer2005}. 

\begin{figure*}  
\begin{center}   
\psfig{figure=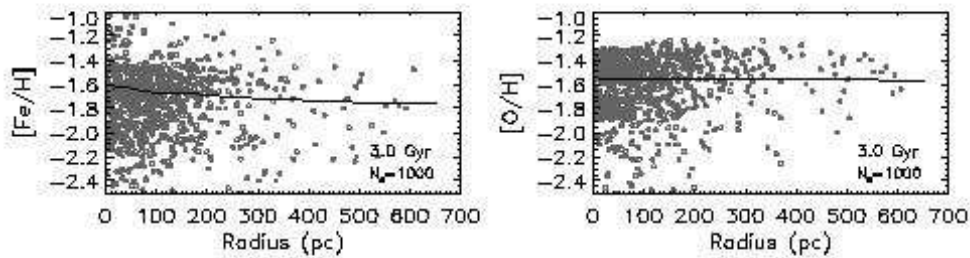,width=0.9\textwidth}   
\end{center}   
\caption{Radial distribution of [Fe/H] (left panel) and [O/H] (right 
panel) at the end of the simulation for the model Draco-25.  The 
solid lines represent the radial profile of the mean values.  $N_{\rm 
s}$ has the same meaning as in Fig. \ref{fig:otofe}.} 
\label{fig:zraggio25}   
\end{figure*}   
   
\item One zone chemical models for Draco (and other dSphs) have been 
presented also by IA who do not take into account the possibility of 
galactic winds and assume that the gas removal should eventually 
result from an external mechanism, just as in our scenario.  In 
absence of a wind, the IA interpretation of the sudden decline of 
[$\alpha$/Fe] in the [$\alpha$/Fe]-[Fe/H] diagram is based on the 
different temporal behaviours of the SN II and SN Ia rates. Given the 
much longer evolutionary timescales of these latter, an (assumed) 
sudden switchover of the iron source from SNe II to SNe Ia creates the 
break in the [$\alpha$/Fe] ratio at a certain [Fe/H] value, 
corresponding to the time $t_{\rm Ia}$ at which a significant number 
of SNe Ia is assumed to explode. This time was evaluated to be 
$t_{\rm Ia}\sim $ 1.5 Gyr by \citet{yoshii1996} who studied the stars 
in the solar neighbourhood.  Before $t_{\rm Ia}$ the effect of SNe Ia 
on the metallicity is not considered; for this reason the plateau 
obtained by IA is totally flat (see their Fig. 1). After 1.5 Gyr the 
Fe ejected by SNe Ia produces a decline in [$\alpha$ /Fe]. However, 
this decline is less pronounced than in our models and in those of 
LM. Indeed, such a decrease must not be confused with the decreasing 
branches obtained by the models of LM and ours. Rather, it is 
equivalent to the slow decrease of the plateau shown in these latter 
models where the action of the SNIa starts (and grows) just after 30 
Myr. As a check, we calculated the average ratios [O/Fe] and [Fe/H] 
inside the stellar volume mimicking an one zone model. The results are 
shown in the lowest panels of Fig.~\ref{fig:otofe}: no break is 
present and the plateau has a slope similar to the mean slope of IA. 
In order to take into account the larger decrement of [$\alpha$/Fe] 
observed IA must invoke a stellar mass function steeper than the 
Salpeter one, an assumption not necessary in our models. 
 
\item Cosmological simulations of the dynamical and chemical evolution 
of dSphs have been recently performed by RG. These authors point out 
that objects with masses below $\sim 10^8$ $\rm M_{\odot}$ are able to 
form stars before the reionization epoch (12.5-13 Gyr ago) losing 
their gas because of the photoheating by massive stars inside the 
galaxy (and not by the external ionizing background). More massive 
objects, instead, retain a fraction of their gas even after the 
reionization. Thus, the duration of the star formation in small 
objects is rather short ($< 1$ Gyr), while in more massive galaxies it 
can be protracted for a longer time. RG identify these latter galaxies 
with the observed dwarf irregulars, while the observed dSphs 
properties are reproduced by the smaller simulated objects. These 
latter objects show a metallicity spread that can be understood in 
terms of hierarchical accretion of subhalos containing stars with 
different metallicities. RG thus argue that the observed metallicity 
spreads are not necessarily an indication of star formation extended 
over several Gyr, as claimed by \citet{grebel2004}. 
 
Comparing the above results with ours, we point out that the longer 
star formation assumed in our models (with $M_{\rm h}=7\times 10^8$ 
M$_{\odot}$) is marginally consistent with the finding of RG. We also 
stress that, although the metallicity spreads found in our models 
derive from a completely different mechanism, yet we confirm that 
large spreads of [Fe/H] can rise quite early in our simulations 
(cf. Fig. 7).

\begin{figure}   
\begin{center}   
\psfig{figure=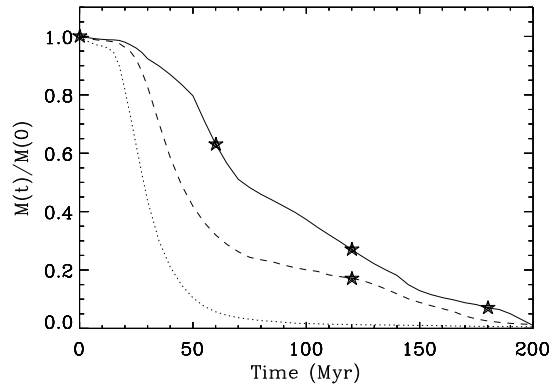,width=0.45\textwidth}   
\end{center}   
\caption{Time evolution of the mass of the cold ISM ($T < 2 \times 
10^4$ K) inside the galactic region for the Draco-S models. The mass 
is normalized to the initial mass inside the galactic region. Solid 
line: Draco-50; dashed line: Draco-25; dotted line: Draco-10. The 
stars indicate the time at which the stellar bursts occur.} 
\label{fig:massa80}  
\end{figure}

\begin{figure}   
\begin{center}   
\psfig{figure=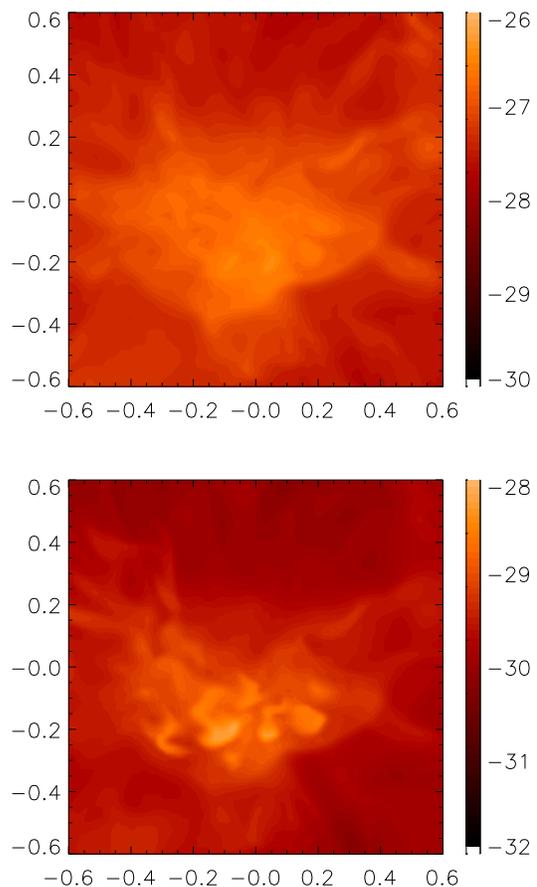,width=0.35\textwidth}   
\end{center}   
\caption{Central distribution on the $z=0$ plane of the SN II (upper 
panel) and SN Ia (lower panel) ejecta densities for the Draco-25 model 
at $t=1520$ Myr. These panels are a close up view of the last two 
panels of the fourth column in Fig~\ref{fig:draco25}. Distances are 
given in kpc.} 
\label{fig:zoom}   
\end{figure}

On the other hand it is known \citep{shetrone2001} that dSphs have 
[$\alpha$/Fe] ratios $\sim0.2$ dex lower than those of the Galactic 
halo fields stars in the same [Fe/H] range, whose [$\alpha$/Fe] ratios 
are 0.3 \citep{shetrone2001}. This usually is interpreted as a 
signature of a longer star formation occurred in dSphs in order to 
allow a larger production of iron by SNe Ia \citep{matteucci2003}.  
Alternative suggestions based on different IMF \citep[e.g.][]{shetrone2001} 
or on less rotation of massive stars (compared to solar neighbourhood) 
have also been invoked \citep{Tsujimoto2006}, but we do 
not consider here these speculative hypotheses. 
 
A glance at Fig.~\ref{fig:otofe} illustrates that in our simulations a 
decrement of $\sim 0.2$ dex in the plateau is achieved after $t\sim 
2.0$ Gyr.  Thus we conclude that, at least in our models, only a 
prolonged star formation history ($> 2.0$ Gyr) can account for the 
chemical differences between the Galactic halo and the dSphs. Such a 
long SFH for dSphs is also found in the chemical models by 
\citet{fenner2006} who are unable to reproduce the Ba/Y ratio unless 
stars formed over an interval long enough for the low-mass stars to 
pollute the ISM with $s$-elements. 
 
Fig. \ref{fig:otofe} also shows that the formation at early times of
stars with low [O/Fe] and relatively large [Fe/H] is not precluded,
although must be regarded as a rather rare event.  Thus our models do
not exclude short star formation durations, although the formation of
stars with low [$\alpha$ /Fe] is less probable.
 
\item Cosmological simulations of dSphs have been worked out also by 
\citet{kawata2005} who, in line with RG, find a rather short ($< 1$ 
Gyr) duration of the star formation. \citet{kawata2005} focus on 
Sculptor galaxy in which Tolstoy et al. (2004) found two distinct 
populations, with the lower metallicity stars more spatially extended 
than the higher metallicity stars. This is interpreted by 
\citet{kawata2005} as a consequence of the metallicity gradient 
which is present in their simulated system. This gradient is similar 
to those found in our simulations at early times; in our models, 
however, the gradients decrease with time. We thus find that, if two 
distinct stellar populations arise from a metallicity gradient 
(different possible mechanisms leading to different populations are 
given by Tolstoy et al. 2004), then their presence is indicative of a 
short star formation history. 
 
We conclude our comparison with the model by \citet{kawata2005} 
briefly highlighting two differences: $i$) our models do not get rid 
of the gas, while the in model by \citet{kawata2005} the SN feedback 
leads to a blow-out phase; $ii$) no break is present in the 
[O/Fe]-[Fe/H] diagram obtained by \citet{kawata2005}. The first 
difference is due to the initial amount of gas assumed by these 
authors, which is an order of magnitude lower than ours; as a 
consequence, the gas binding energy is much lower and the radiative 
losses (which are proportional to the square of the gas density) are 
greatly reduced, allowing the SNe to vent the gas away. The second 
difference comes, as fully realized by \citet{kawata2005}, from having 
these authors assumed for the SNe Ia the model proposed by 
\citet{kobayashi2000} who suggested that SNe Ia are inhibited in the 
stars with [Fe/H]$<-1$. 
 
\end{itemize} 
  
\section{Discussion and conclusion}  
\label{sec:discussion} 
 
\subsection{Dynamical evolution}  
  
We run several 3D numerical models in order to understand the chemical 
and dynamical evolution of the ISM in dSphs as a function of several 
parameters like the SFH and the amount of dark matter. Given the large 
amount of cpu time required to complete a single model, we could not 
explore a vast region of the parameter space; instead, we tailored our 
galaxy model on Draco and considered SFHs similar to those supported 
by the observations for this galaxy. All the adopted SFHs last 3 Gyr 
and give rise to the same final amount $M_{\star}=5.6\times 10^5$ 
M$_{\odot}$ of stellar mass.  The stars are supposed to form in a 
sequence of instantaneous identical bursts separated by quiescent 
periods; different SFHs differ by the assumed number of bursts, but 
the total number of SNII (but not SNIa) is always the same. 
  
Although the total energy released by the SNe is orders of magnitude
larger than the binding energy of the ISM, the galaxy retains almost
the totality of its initial gas, unless rather light dark haloes are
assumed. This is due to the huge efficiency of the radiative cooling,
despite the low mean metallicity ($Z\sim 10^{-2}$ $\rm Z_{\odot}$) of
the gas. For most parameter combinations tested in this study, such
effective radiative losses prevent the gas to slowly accumulate the SN
energy. With our assumptions, the galaxy never gets rid of its gas,
unless the potential well is rather shallow (see
Section~\ref{sec:draco-s}); in this latter case a galactic wind starts
quite soon and the galaxy looses all its gas in less than 200 Myr.
  
In the light of the above discussion, we conclude that the dark halo 
of Draco must be massive and extended in order to retain the gas for a 
period of several Gyr, the duration of its star formation.  This in 
turn implies the need of an external mechanism to remove the gas and 
end the star formation, as gas stripping and/or tidal interaction by 
the Galaxy \citep{mayer2005}. 
  
\subsection{Radiative losses}  
 
The above arguments highlight the crucial role played by the radiative 
cooling in our models: would these losses be less substantial, 
galactic winds could develop more easily and the evolutionary scenario 
would drastically change. A correct evaluation of the radiative losses 
in the numerical simulations is thus of the utmost 
importance. Unfortunately, it is known that in numerical hydrodynamics 
several factors concur to degrade an accurate estimate of these 
losses. More confusing, some factors lead to an overestimate, while 
others to an underestimate. 
 
Overestimates occur particularly at the contact discontinuities 
separating hot rarefied and cold dense gas phases because the 
intrinsic diffusion of the numerical scheme spreads these 
discontinuities over several mesh points creating a gas phase with 
intermediate densities and temperatures characterized by a large 
emissivity. Although several physical processes (such as turbulences 
and heat conduction) really smear out these discontinuities, the 
numerical spread is likely to be larger than the physical one. 
 
Underestimates as well as overestimates of the radiative cooling at 
the contact discontinuities may occur when in the numerical code a 
cooling curve calculated in a regime of ionization equilibrium is 
implemented (as we actually do). A hot plasma undergoing strong 
radiative losses may cool at lower rates \citep{sutherland1993} or at 
higher rates \citep[e.g.][]{borkowski1990} relative to the condition 
of ionization equilibrium, depending on its thermal history. 
 
In order to test the sensitivity of our models to the amount of
radiative losses, we rerun model Draco-50 reducing the radiative
cooling by a factor of ten.  Also in this case the great majority of
the gas remains bounded to the galaxy. In particular $\sim 70$\% of
the gas remains inside the galactic volume, while only 30 \% (with
some oscillations) stays in the stellar region. Given the reduced
effectiveness of the radiative cooling, the hot gas can expand more
efficiently pushing most of the ISM at larger distances and creating a
bubble encompassing a large fraction of the stellar volume. We thus
conclude that the ability to retain the gas is a robust property of
our model and depends essentially on the prolonged and thus low SNe II
rate. As an argument in favour of this conclusion, we quote
\citet{mori2002} who worked out a 3D simulation of a subgalactic
object with a potential well and an amount of ISM quite similar to
our, with a gas binding energy four times larger but a number of SNe
II only two time higher. Despite this more disadvantageous proportion,
most of the gas is blown away because the SNe II are made exploding
all together instantaneously.
 
\subsection{Chemical evolution} 
 
Although the majority of the metals expelled by SNe is still inside
the dark matter halo or in its neighbourhood (and is still bounded to
the galaxy), only a small fraction ($\sim 18\%$) is located inside the
stellar region. This is the amount of metals which contributes to the
metallicity of the forming stars.  Thus, being the chemical elements
observed in stars the main observational constraints in chemical
evolution models, assuming that the majority of the SN products stay
in the observable zone may lead to infer non realistic galactic
properties and/or stellar properties. No matter if metals are
lost by galactic winds \citep[as assumed by some dSphs chemical models][]
{robertson2005,carigi2002} whether are pushed at higher distances, our
finding is constitent with the assumption that the majority of the
metals do not enrich the newly forming stars.
 
At the early stage of the evolution the [Fe/H] distribution in the ISM
is rather spread.  As the number of the bursts increases with time,
the metals expelled by SNe become distributed more and more
homogeneously. This behaviour is reflected in the MDF which narrows
around the mean value $\langle$[Fe/H]$\rangle$, which increases with
time. At the end of the simulations we obtain a mean value of
$\langle$[Fe/H]$\rangle \sim$-1.65 with a spread of $\sim$ 1.5 dex
\citep{lehnert1992, shetrone2001, aparicio2001, bellazzini2002}, while
the maximum value of the distribution occurs at [Fe/H]=-1.5 for the
reference model Draco-50, and at [Fe/H]=-1.6 for Draco-25.
 
We can also reproduce satisfactory the general observed behaviour of 
[O/Fe] in the [O/Fe]-[Fe/H] diagram. In dSphs this diagram shows a 
break in [O/Fe] occurring at [Fe/H]$\sim -1.6$, with an approximate 
constant level of [O/Fe] at lower metallicities and a monotonic 
decrease at larger metallicities.  In agreement with previous chemical 
evolution models (e.g. LM and IA) we regard the stars populating the 
plateau as polluted mainly by metals produced by SNe II. The shallow 
decline of the plateau at larger values of [Fe/H] is due to the 
contribution to the iron enrichment by SNe Ia which increases with 
time. 
 
Instead, we attribute the decreasing branch in the [O/Fe]-[Fe/H]
diagram to the low value of the porosity of SNIa remnants: given the
low SNIa rate, these remnants are located quite apart one from
another, and the iron ejected by SNe Ia is distributed rather
inhomogeneously through the stellar volume
(cf. Fig.~\ref{fig:zoom}). As a consequence, stars forming in the
(relatively small) volume occupied by SNIa remnants have a ratio
[O/Fe] lower than those forming elsewhere. This effect can naturally
account for the stars with low [O/Fe] and high [Fe/H], without
necessarily invoking a complex SFH as suggested for some dSphs
\citep{carigi2002}.
 
In our models the plateau and the descending branch become populated 
at the same time, although with different proportions. As the galaxy 
evolves, the break moves toward larger values of [Fe/H]. This 
interpretation of the [O/Fe]-[Fe/H] diagram differs from previous 
chemical models in which the break appears only after a specific time, 
without changing its position later on. LM interpret the break as a 
sign of the occurrence of a galactic wind. IA, who adopt a closed-box 
model, associate the break to a sudden switchover of the SNe Ia 
explosions at a specified time $t_{\rm Ia}\sim 1.5$ Gyr. We claim that 
there is no direct link between the break presence (and position) and 
$t_{\rm Ia}$. This conclusion is reinforced by the time evolution of 
the SNIa rate (cf. the solid line in Fig.~\ref{fig:snirate}) which, 
after a rapid rise before $\sim 0.5$ Gyr, varies only by a factor of 2 
over 2.5 Gyr; this smooth temporal profile can not justify the break 
in the [$\alpha$/Fe]-[Fe/H] diagram (unless a drastic reduction in the 
star formation rate occurs). 
 
In the end, one possible discrimination between the LM scenario and the  
our would be given by the possible observation of some stars located  
"below" the plateau in the [$\alpha$/Fe]-[Fe/H] diagram, i.e.  
showing a [O/Fe] ratio lower than that of the plateau at low values 
of [Fe/H]. 
 
We also stress that observations of ``transition'' galaxies 
\citep{grebel2003} which have not yet lost all their gas content may 
discriminate between the LM scenario and the our. In the former one, 
infact, these galaxies are expected to show only the plateau in the 
[O/Fe] distribution, while in the latter one the descending branch 
would be already present at low values of [Fe/H].

\subsection{Summary} 
 
In conclusion, our major findings are the following: 
 
1) The long duration of the star formation attributed to Draco (and 
the other local dSphs too), together with the high effectiveness of 
the radiative losses, exclude that the galaxy got rid of its gas by an 
internal mechanism such as a galactic wind. The gas removal to 
complete star formation and to evolve to a gas-poor system should 
result from external mechanisms such as ram pressure stripping and/or 
tidal interaction with the Galaxy. 
 
2) Although the SN ejecta remain gravitationally bounded during the
star formation, yet only a low fraction ($\sim 18$\%) stays in the
region where star forms. This effect should be taken into account in
chemical evolution models.
  
3) Our models succeed in reproducing the [Fe/H] distribution function 
of the stars. In agreement with observations, we find a mean value 
$\langle$[Fe/H]$\rangle = -1.65$ with a spread of $\sim 1.5$ dex. 
  
4) We can also satisfactory reproduce the observed [O/Fe] vs [Fe/H] 
diagram. 
 
5) Contrary to the usual interpretation, we rule out any particular 
relation between the [$\alpha$/Fe] break position in the 
[$\alpha$/Fe]-[Fe/H] plane and the onset of SN Ia explosions. Methods 
intended to determine $t_{\rm Ia}$ in this way might be incorrect. 
 
6) In agreement with observations, our models develop initially 
moderate metallicity gradients which become weaker and weaker as the 
galaxy evolves. 
 
7) Finally, the above results are slightly dependent on the particular 
star formation history adopted.

\section*{Acknowledgements} 
 
We thank the referee, Leticia Carigi, whose comments greatly improved
the presentation of the paper.  We are very grateful to M. Bellazzini
for giving us the data used in our Fig.~\ref{fig:zstellenosni}. We are
also indebted to G. Lanfranchi and F. Matteucci to sent us some
simulations in advance of publication to compare with our models. Many
thanks also to L. Mayer and all the above friends for useful
suggestions and discussions which greatly improved the final
version. We acknowledge financial support from National Institute for
Astrophysics (INAF). The simulations were run at the CINECA
Supercomputing Centre with CPU time provided by a grant of the
National institute for Astrophysics (INAF). S.R. acknowledges
financial support from the Deutsche Forschungsgemeinschaft (DFG) under
grant TH 511/8

\bibliographystyle{mn2e} 
\bibliography{dwarf_refs} 
 
\appendix 
\section{} 
 
A SNR expanding in a homogeneous uniform medium evolves initially 
following the Sedov solution during which radiative losses can be 
neglected. With time, however, such losses eventually cause the formation
of a dense shell, and the high pressure of the hot interior pushes this 
shell (the snowplow phase).  Assuming an explosion energy of 10$^{51}$ 
erg, the time at which the Sedov phase ends is given by 
\citet{cioffi1991} $t_0=1.49\times 
10^4n_0^{-4/7}\zeta^{-5/14}$ yr, where $n_0$ is the density of the 
ambient medium and $\zeta$ is the ISM metal abundance in solar units. 
The SNR stalls when its expansion velocity drops to the 
the local sound speed $c_0$; the stalling radius is given by 
\citet{cioffi1991} $R_{\rm st}=59n_0^{-6/35}P_4^{-1/5}\zeta ^{-2/35}$ 
pc, in terms of the pressure of the ambient medium $P_4=P/(k_{\rm 
B}\times 10^4\;{\rm cm^{-3}\;\rm K})$ (where $k_{\rm B}$ is the 
Boltzman constant), after a time $t_{\rm st}=1.38n_0^{0.33}P_4^{-7/10} 
\zeta ^{-2/35}$ Myr.  Eventually, the SNR cavity is refilled by the 
ISM on a timescale of the order of $t_{\rm fill}=R_{\rm st}/c_0$. 
 
It is known that a single SNR does not contribute significantly to the 
ISM energization; at the end of its evolution only a few percent of 
the explosion energy has been fed into the surrounding gas, while the 
rest has been radiated away \citep[e.g.][]{bradamante1998}.  However, 
when many SNe explode with a rate per unit volume $S$ high enough, 
their combined effect may lead to the thermalization of a very high 
fraction (close to 100\%) of the explosion energy. In fact, if many 
SNRs collide with each other the shells fragment and the hot gas of 
their interior merges forming a dilute medium. Repeated supernova 
explosions are especially effective in reheating this medium because 
the SNR cooling time is relatively long in a low density region. A 
considerable fraction of gas may thus be driven out of the galaxy. To 
understand when this is effectively the case, one has to calculate the 
collision time $t_{\rm int}$. This time can be defined as the time 
required for SNRs forming at the rate $S$ to fill up the volume of 
space.  If the expansion law of a remnant is given by $R_{\rm 
s}=At^{\alpha}$ (with $A$ and $\alpha$ constants), then \citep[cf.][] 
{larson1974, melioli2004} $t_{\rm int}=({3 \alpha+1 
\over SA^3})^{1/(3 \alpha +1)}$.  Unless for very low values of $n_0$, 
the Sedov phase represents only a short stage in the evolution of a 
SNR whose bulk expansion can be described by the snowplow law. In this 
case we have $R_{\rm s}=1.35n_0^{-13/49}\zeta^{-2/49}t^{2/7}_{\rm yr}$ 
pc, where $t_{\rm yr}$ is the time in years \citep{cioffi1991}. 
 
In order to make some quantitative consideration concerning our 
standard model, we assume $n_0=0.1$ cm$^{-3}$, $P_4=0.1$, ${\cal 
R_{\rm SN II}}=3.7\times 10^{-6}$ yr$^{-1}$ and a mean value 
$S=3.28\times 10^{-15}$ pc$^{-3}$ yr$^{-1}$.  Then we have $t_{\rm 
int}=20.1 \zeta^{0.07}$ Myr, which is longer than $t_{\rm st}$. This 
means that the SNRs do not interact one with another during their 
evolution because they stop their expansion before colliding. 
Despite this result, with a sufficiently high rate $S$, and a long 
lifetime $t_{\rm fill}$, the SNR cavities would still necessary 
overlap, connect, and form a network of tunnels.  One can characterize 
this topology in terms of the porosity parameter $Q=SV_{\rm st}t_{\rm 
fill}$, where $V_{\rm st}=(4 \pi /3)R^3_{\rm st}$. The condition $Q>1$ 
implies that SNRs overlap. With our assumptions we get $Q=0.5\zeta 
^{-8/35}$. We thus conclude that in our model a combined action of 
SNRs pressurizing the ISM can be obtained for $\zeta<0.05$. Assuming 
$\zeta=0.01$ we calculate that SNRe II fill most of the stellar volume 
after $t\sim 12$ Myr. 
 
\label{lastpage} 
\end{document}